\documentclass[12pt]{iopart}

\usepackage{iopams}

\usepackage{mathrsfs}

\expandafter\let\csname equation*\endcsname\relax
\expandafter\let\csname endequation*\endcsname\relax

\usepackage{caption}
\usepackage[hidelinks]{hyperref}
\usepackage{subcaption}
\usepackage{graphicx}               
\usepackage[utf8]{inputenc}         
\usepackage{xstring}
\usepackage{soul,todonotes}
\usepackage{cite}
\usepackage[noabbrev]{cleveref}
\crefformat{equation}{equation~#2#1#3}

\newcommand{\kindex}[2]{\ensuremath{{#1}_{\scalebox{0.5}{#2}}}}
\newcommand{\U}{\textrm{U}}
\newcommand{\Uinf}{\kindex{\U}{$\infty$}}
\usepackage[locale=UK,
number-mode=math,
unit-mode=text,
per-mode=symbol,
number-unit-product=\ ,
retain-zero-exponent=false]{siunitx}[=v2]
\DeclareSIUnit{\pixel}{px}
\DeclareSIUnit{\fps}{fps}

\usepackage{etoolbox} 
\robustify{\kindex}

\graphicspath{{./figs/}{../figurematter/latex/figs/}{../figurematter/plots/}}

\begin{document}
	\title[Aeroelastic characterisation of a bio-inspired flapping membrane wing]{Aeroelastic characterisation of a bio-inspired flapping membrane wing}
	\author{Alexander Gehrke, Jules Richeux\footnote{current affiliation: ENSTA Bretagne, 2 rue François Verny, 29806 Brest Cedex 9, France}, Esra Uksul\footnote{current affiliation: TU Delft, Leeghwaterstraat 39, 2628 CB Delft, Netherlands}, and Karen Mulleners}
	\address{École polytechnique fédérale de Lausanne, Institute of Mechanical Engineering,
		Unsteady flow diagnostics laboratory, 1015 Lausanne, Switzerland}
	\ead{karen.mulleners@epfl.ch}
	\vspace{10pt}

	\begin{abstract}
		Natural fliers like bats exploit the complex fluid-structure interaction between their flexible membrane wings and the air with great ease.
		Yet, replicating and scaling the balance between the structural and fluid-dynamical parameters of unsteady membrane wings for engineering applications remains challenging.
		In this study, we introduce a novel bio-inspired membrane wing design and systematically investigate the fluid-structure interactions of flapping membrane wings.
		The membrane wing can passively camber and its leading and trailing edges rotate with respect to the stroke plane.
		We find optimal combinations of the membrane properties and flapping kinematics that out-perform their rigid counterparts both in terms of increased stroke-average lift and efficiency, but the improvements are not persistent over the entire input parameter space.
		The lift and efficiency optima occur at different angles of attack and effective membrane stiffnesses which we characterise with the aeroelastic number.
		At optimal aeroelastic numbers, the membrane has a moderate camber between \SI{15}{\percent} and \SI{20}{\percent} and its leading and trailing edges align favourably with the flow.
		Higher camber at lower aeroelastic numbers leads to reduced aerodynamic performance due to negative angles of attack at the leading edge and an over-rotation of the trailing edge.
		Most of the performance gain of the membrane wings with respect to rigid wings is achieved in the second half of the stroke when the wing is decelerating.
		The stroke-maximum camber is reached around mid-stroke but is sustained during most of the remainder of the stroke which leads to an increase in lift and a reduction in power.
		Our results show that combining the effect of variable stiffness and angle of attack variation can significantly enhance the aerodynamic performance of membrane wings and has the potential to improve the control capabilities of micro air vehicles.
	\end{abstract}
	\submitto{\BB}


	\section{Introduction}%
	Many swimming or flying animals use flexible appendages like fins, tails, and wings for locomotion on land, in water, or in air~\cite{wootton_functional_1992, swartz_mechanical_1996, tytell_hydrodynamics_2004, fish_passive_2006, young_details_2009}.
	Natural fliers have flexible wings which turn out to be an efficient means of propulsion that is robust to perturbations~\cite{boerma_wings_2019}, lightweight~\cite{swartz_wing_1992}, reconfigurable to enhance lift production or reduce drag~\cite{rayner_bounding_1985, lentink_how_2007}, and silent~\cite{jaworski_aeroacoustics_2020}.
	By actively controlling their wing's kinematics, flying animals can alternate between flapping wing flight and gliding to conserve energy~\cite{muijres_aerodynamic_2012, henningsson_efficiency_2014}.
	In addition to the aerodynamic benefits, the wing's elasticity helps to absorb kinetic energy from collisions~\cite{mountcastle_biomechanical_2013}, and some species can store their wings to protect them when moving through tight spaces on the ground~\cite{sun_structure_2012}.\\
	Flying and gliding mammals, in particular several species of bats, lemurs, and squirrels, use membrane wings to create lift.
	Bats have a lightweight bone and muscle structure that allows them to precisely control large shape deformations of their thin and compliant membranes.
	By flapping and deforming their wings, bats can adapt to different flow conditions and perform agile flight manoeuvres~\cite{tian_direct_2006, bahlman_design_2013, muijres_leading_2014}.\\
	Membrane wings have an enhanced aerodynamic performance compared to rigid wings due to their ability to passively camber and reduce their effective angle of attack.
	The membrane deformation leads to higher lift slopes, a delay in stall onset, improved gust alleviation, and enhanced flight stability relative to rigid wings~\cite{song_aeromechanics_2008, jaworski_high-order_2012}.
	These properties make membrane wings a promising example for advances in the design of micro air vehicles~\cite{shyy_flapping_1999, abdulrahim_flight_2005, pines_challenges_2006, stanford_fixed_2008, shyy_recent_2010, Bleischwitz2015, ramezani_biomimetic_2017}.\\
	Even though our natural examples exploit the complex fluid-structure interaction between their flexible or membrane wings and the surrounding air with great ease, it is not trivial to replicate and scale the balance between the structural and fluid dynamical parameters for engineering applications.
	Many theoretical, numerical, and experimental studies have been conducted using flexible and deformable wings to improve our understanding of the impact of fluid-structure interactions on the aerodynamic performance of fixed and unsteadily flapping wings and to provide guidance for the design of micro air vehicles.\\
	Membrane wings at fixed angles of attack have been studied extensively in the past in the context of sailboats, hang gliders, and Cretan type wind mill designs~\cite{thwaites_aerodynamic_1961, nielsen_theory_1963, cook_modelling_2005, fleming_design_1982}.
	First studies on inextensible membranes were focussed on predicting the shape and aerodynamic performance of sails.
	Thwaites~\cite{thwaites_aerodynamic_1961} and Nielsen~\cite{nielsen_theory_1963} were among the first to investigate the relation between tension, aerodynamic loads, and deflection of inextensible sails.
	They found that the equilibrium shape of a passively deformable sail at low angles of incidence $\alpha$ is solely governed by the ratio of the aerodynamic pressure and the tension parameter $\mathscr{T}$ denoted by $\lambda$ \cite{thwaites_aerodynamic_1961}:
	\begin{equation}
		\lambda = 2 \rho \Uinf^2 c / \mathscr{T}\quad.
	\end{equation}
	Different models and analytical solutions for inextensible and flexible membranes have since been proposed building upon the early work of Thwaites and Nielsen~\cite{thwaites_aerodynamic_1961, nielsen_theory_1963}.
	An extensive summary of these works can be found in~\cite{newman_aerodynamic_1987}.\\
	At higher angles of incidence and for larger excess lengths of the membranes, classical, potential flow based theories fail to predict the equilibrium states of fixed membrane wings and numerical studies coupling the Navier-Stokes equations with an aeroelastic model are desirable.
	Based on the results of such coupled simulations, Smith and Shyy~\cite{smith_computational_1995, smith_computation_1996} derived two non-dimensional parameters to characterise the interplay between aeroelastic and aerodynamic effects:
	\begin{equation}
		\kindex{\Pi}{1} = \left(\frac{E h}{\frac{1}{2} \rho U^2_\infty c}\right)^{1/3} \quad \textnormal{and} \quad \kindex{\Pi}{2} = \left(\frac{\kindex{\epsilon}{0} h}{\frac{1}{2} \rho U^2_\infty c}\right)^{1/3} \, ,
		\label{eq:effective_stiffness}
	\end{equation}
	with $E$ the Young's modulus of the membrane material, $h$ the membrane thickness, $\rho$ the fluid density, $\Uinf$ the freestream velocity, $c$ the chord length, and \kindex{\epsilon}{0} the membrane pretension.
	The two parameters \kindex{\Pi}{1} and \kindex{\Pi}{2} control the steady-state, inviscid aeroelastic behaviour of an initially flat membrane at fixed angles of attack.
	The dimensionless deformation of the membrane wing is inversely proportional to \kindex{\Pi}{1} in the absence of pretension and inversely proportional to \kindex{\Pi}{2} in the limit of vanishing material stiffness~\cite{shyy_aerodynamics_2008}.\\
	If the load distribution on the membrane wing is assumed to be uniform, the maximum camber \kindex{z}{max} is entirely characterised by the Weber number \textit{We} and the membrane pretension \kindex{\epsilon}{0}~\cite{song_aeromechanics_2008}.
	The Weber number is defined as the ratio between the aerodynamic loading and the effective stiffness of the membrane:
	\begin{equation}
		\textit{We} = \kindex{C}{L} \frac{\frac{1}{2} \rho \Uinf^2 c}{E h} =  \frac{\kindex{C}{L}}{\textit{Ae}}\, ,
		\label{eq:weber_number}
	\end{equation}
	where \textit{Ae} is defined as the aeroelastic number~\cite{waldman_camber_2017}:
	\begin{equation}
		\textit{Ae} = \frac{E h}{\frac{1}{2} \rho U^2_\infty c} \,,
		\label{eq:aeroelastic_number}
	\end{equation}
	and relates to the effective stiffness introduced in \cref{eq:effective_stiffness} as $\textit{Ae} = \kindex{\Pi}{1}^3$~\cite{smith_computational_1995, waldman_camber_2017}.\\
	The membrane shape characterisation by~\cite{song_aeromechanics_2008} was extended by Waldman and Breuer~\cite{waldman_camber_2017} who included the Young-Laplace equation for non-linear deformation of the membrane at low angles of attack, assumed a uniform pressure distribution, and incorporated a potential flow model to estimate the aerodynamic loading from the thin membrane airfoil.
	Their model shows a remarkable agreement with experimental data of the maximum camber of membranes for different material properties, fluid characteristics, and wing angles of attack expressed in terms of the aeroelastic parameters \textit{Ae} and \textit{We}.
	By coupling thin airfoil theory with membrane equations, a new analytical model was derived by Alon Tzezana and Breuer~\cite{alon_tzezana_thrust_2019} to predict the membrane shape even more accurately than \cite{waldman_camber_2017} for steady and unsteady membrane wings under various conditions.\\
  %
  The deformation of flexible membrane aerofoils typically manifests in the form of a positive wing camber.
  The positive camber leads to an increase in the lift coefficient, a decrease in the drag coefficient, and an increase in the lift-to-drag ratio for flexible wings compared to their rigid counterparts especially at angles of attack close to the stall angle of the wing~\cite{song_aeromechanics_2008, hu_flexible-membrane_2008}.
  The curvature of the leading edge of a cambered membrane allows the flow to stay attached at higher angles of attack~\cite{koekkoek_stroke_2012}.
  The rigid wings at the same angles of attack have larger flow separation regions and are subjected to large-scale vortex shedding accompanied by a loss in lift and an increase in drag~\cite{rojratsirikul_unsteady_2009, rojratsirikul_flow-induced_2011}.
  The fluid-structure interaction between the shed vortices and the membrane leads to a reduction in the separation area and lower surface pressure on the wing below the separation area.
  These changes in flow topology lead to an increase in lift at moderate angles of attack ($\alpha = \ang{10} - \ang{25}$), but the unsteady membrane fluctuations cause a loss in lift-to-drag ratio at the lower angles of attack~\cite{gordnier_high_2009, rojratsirikul_effect_2010, gordnier_impact_2014}.
  The dominant membrane vibrations are coupled with the shedding frequency of large-scale flow structures \cite{he_fluidstructure_2020}.
  Membrane wings with lower aspect ratios ($\approx\num{1}$) experience higher frequency vibrations and higher vibration mode shapes due to an increase in downwash and a delay of vortex shedding to higher angles of attack~\cite{Bleischwitz2015}.
  The average membrane shape is not very sensitive to changes in angle of attack~\cite{rojratsirikul_unsteady_2009}.\\
	In many technical application, the wing kinematics or the flow around flexible wings is highly unsteady and vortex dominated, which further complicates the fluid-structure interactions.
	Due to unsteady flow conditions and wing kinematics, flapping wings are subject to highly unsteady aerodynamic loadings within each wing stroke~\cite{sane_control_2001} and the flow development is governed by the formation and shedding of large-scale coherent leading edge vortices~\cite{birch_force_2004, shyy_flapping_2007, eldredge_leading-edge_2019}.
	The large inertial force variations on the wing give rise to unsteady and non-linear wing deformations linking the analysis of fluid and structural dynamics on the flexible wings inevitably together.
	To explore the shape and force response of membrane wings subjected to an unsteady flow or unsteady wing kinematics, several experimental~\cite{tregidgo_unsteady_2013} and numerical~\cite{gopalakrishnan_effect_2010, jaworski_high-order_2012, jaworski_thrust_2015, sekhar_canonical_2019} studies have been conducted in the recent past.
	Membrane wings have also been applied and tested in novel flapping wing micro aerial vehicle~\cite{hu_experimental_2010, bahlman_design_2013, ramezani_biomimetic_2017} and flapping wing energy harvester designs~\cite{mathai_fluidstructure_2022}.\\
  Similar to the steady response, pitching membrane wings experience reduced and delayed flow separation~\cite{tregidgo_unsteady_2013, gopalakrishnan_effect_2010, mathai_fluidstructure_2022}.
  The flow on the membrane wings remains attached at higher angles of attack and reattaches earlier when pitching down compared to the flow on rigid wings.
	Analysis of the flow topology revealed that the flow field is dominated by the growth of a strong leading edge vortex which eventually lifts off of the wing for rigid flapping wings.
	The leading edge vortex stays bound to the wing and spreads over the entire chord when a highly cambering membrane wing is used~\cite{gopalakrishnan_effect_2010, mathai_fluidstructure_2022}.
	This leads to enhanced force production on the membrane wings compared to the rigid wings.
	The thrust and propulsion efficiency of sinusoidally heaving and flapping membranes was numerically investigated by Jaworski and Gordnier~\cite{jaworski_high-order_2012, jaworski_thrust_2015} with a high-order Navier-Stokes solver coupled to a non-linear membrane structural model.
  The maximum performance is achieved for specific sets of membrane pre-strain \kindex{\epsilon}{0} and elastic modulus $E$.
  The formation and shedding of a large-scale leading edge vortex every half-cycle interact advantageously with the local membrane camber to enhance the propulsive force under the thrust optimal conditions.
  At efficiency optimal aeroelastic conditions, the vortex shedding reduces the magnitude of the unsteady lift and minimises the pressure drag and required power to perform the flapping motion~\cite{jaworski_thrust_2015}.
  The lift and thrust production of a tethered flapping wing vehicle with different membrane flexibilities in forward flight was investigated experimentally by Hu et al.~\cite{hu_experimental_2010}.
  The rigid wings generate more lift at higher advance ratios $J > \num{0.6}$ ($J = \kindex{U}{$\infty$} / (2 f A)$).
  At lower advance ratios $J < \num{0.6}$, which corresponds to increased flapping frequencies $f$, the flexible latex wings produce higher lift.
	The flexible wings produce more thrust than the rigid wings across all frequencies.\\
  These examples reveal a clear potential for flexible membrane wings to out-perform their rigid counterparts but they also indicate that the improvements are not persistent over the entire input parameter space~\cite{sekhar_canonical_2019}.
  Systematic investigation on the influence of unsteady fluid-structure interactions on the aerodynamic performance of flapping membrane wings for a large range of flapping motions and membrane properties is highly desirable to provide further guidance to the design of human-engineered flexible wing fliers~\cite{bomphrey_insect_2018}.\\
	%
	Here, we present a novel bio-inspired membrane wing design with self-cambering and flow alignment capabilities to systematically study fluid-structure interactions related to passively deformable flapping wings.
	We experimentally characterise the force response of membrane wings over a large range of flapping kinematics and wing material properties on a robotic flapping wing mechanism.
	Additional deformation measurements are conducted to capture the membrane shape throughout the full flapping wing cycle.
	We demonstrate that the connection between membrane shape and aerodynamic force production can be explained and scaled by two non-dimensional numbers, the aeroelastic number \textit{Ae} and the Weber number \textit{We}.
	A unique combination of aeroelastic properties and angle of attack is found to achieve either highest lift or most efficient hovering flight.
	Finally, the implications of our findings for the design of micro air vehicles using membrane wings are discussed.
	\section{Methods}%
	\subsection{Membrane wing model}
	\begin{figure}
		\centering
		\includegraphics[]{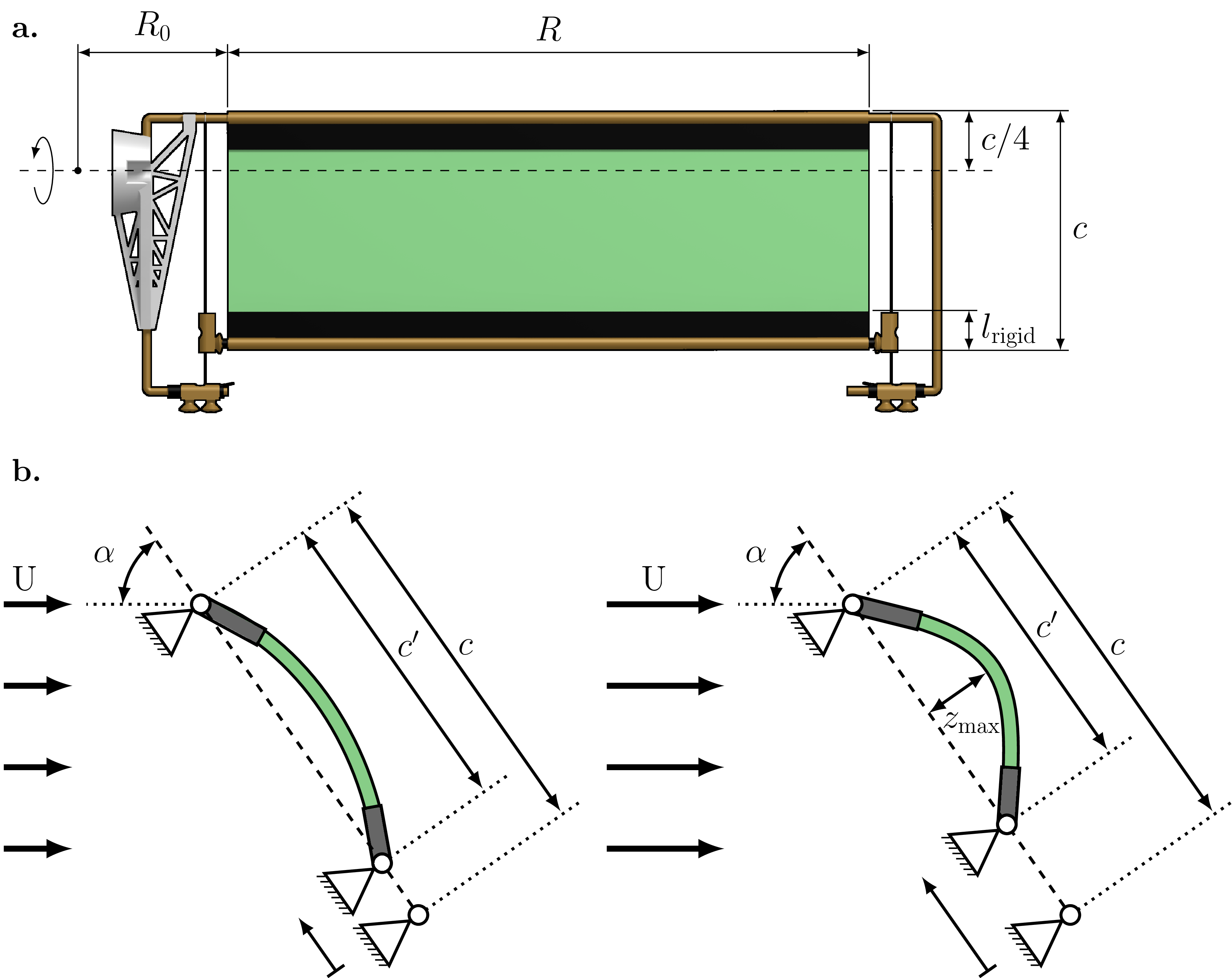}
		\caption{a.~Detailed drawing of the novel membrane wing design. b.~Side view of the wing during loading indicating the passive deformation mechanism. The angle $\alpha$ is the wing’s angle of attack relative to the stroke plane.}
		\label{fig:BillyBoy}
	\end{figure}%
	In this study, we present a novel bio-inspired membrane wing design shown in its flat and undeformed shape in \cref{fig:BillyBoy}a.
	The leading and trailing edges of the wing are rigid and have brass bearings which allow the edges to rotate around their span-wise axes.
	The trailing edge has an additional degree of freedom as it can translate in the chord-wise direction in a frictionless manner.
	No pre-tension is applied to the membrane or the trailing edge sliders.
	Depending on the aerodynamic pressure difference between the pressure and suction side of the membrane, the wing passively cambers and the distance between the leading and trailing edge shortens (\cref{fig:BillyBoy}b).
	The adaptive wing is designed to enhance the aerodynamic performance of the wing in two ways:
	1) the camber of the wing grows with increasing aerodynamic pressure, and 2) the leading and trailing edges of the wing rotate and align favourably with the flow.
	The combined effects of these two mechanisms yield higher lift versus angle of attack slopes, higher maximum values of the lift coefficient, and delayed stall to higher angles of attack due to a reduction in the effective angle of attack at the leading edge.\\
	The compliant membranes are made of a silicone-based vinylpolysiloxane polymer (\textit{Zhermack Elite Double 32 shore A}) created by mixing a base and catalyst in a 1:1 ratio and centrifuging the mixture to homogenise it.
	The mixture is poured into a flat cast to create different silicon sheets of $h =$ \SIrange{0.3}{1.4}{\milli\metre} thickness.
	The membrane sheets have a density of $\rho = \SI{1160}{\kilo\gram\per\metre\cubed}$ and a Young's modulus of $E = \num{1.22} \pm \SI{0.05}{\kilo\pascal}$~\cite{grandgeorge_mechanics_2021}.
	Rectangular wings are cut from the silicon sheets and fixed between two thin carbon fibre plates without pre-stretching the membranes (\cref{fig:BillyBoy}a).
	The rigid wings serving as a reference case have a Young's modulus more than three orders of magnitude higher than the membrane wings $E = \SI{1.33}{\mega\pascal}$ and a thickness of $h = \SI{1}{\milli\metre}$.
	The reinforced membranes are glued to hollow, cylindrical brass rods which serve as the outer race of the leading and trailing edge bearings.
	The wings are mounted on the wing frame which itself is connected to the load cell and the flapping wing mechanism~\cite{gehrke_phenomenology_2021}.
	The membrane wings have a chord length $c = \SI{55}{\milli\metre}$ in their undeformed state and a wing span $R = \SI{150}{\milli\metre}$ (\cref{fig:BillyBoy}a).
	The root cut-out \kindex{R}{0} is the distance from the stroke-rotation axis to the root of the wing and is constant for all experiments.
	The pitch rotation axis is at a quarter chord length from the leading edge.
	The rigid carbon fibre plates and brass bearings make up $\kindex{l}{rigid} = \SI{8.5}{\milli\metre}$ or $\SI{0.15} {c}$ at the leading and trailing edges of the wings.
	A video of the membrane wings in motion and further information about the wing platform can be found in the Gallery of Fluid Motion contribution\footnote{\href{https://gfm.aps.org/meetings/dfd-2021/61290c99199e4c7029f449aa}{V0011: \emph{Don't be rigid, be BILLY}}}~\cite{gehrke_video_2021}.
	\subsection{Wing kinematics and dynamic scaling}
	\begin{figure}
		\centering
		\includegraphics[]{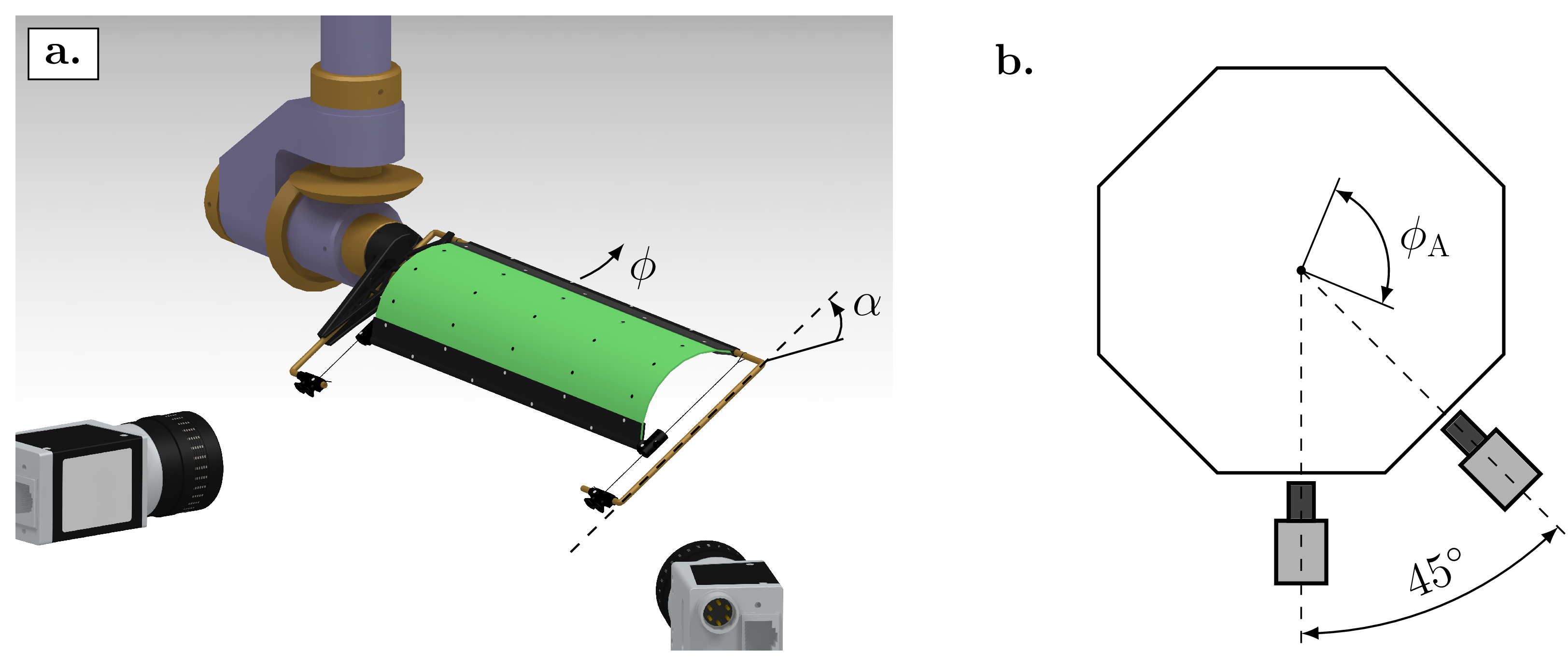}
		\caption{a.~Drawing of the membrane wing mounted on the flapping wing apparatus.
			The angle $\alpha$ is the wing's angle of attack relative to the stroke plane.
		b.~Schematic visualisation of the stereo deformation measurement configuration including the camera positions relative to the peak-to-peak wing stroke amplitude indicated by \kindex{\phi}{A}.}
		\label{fig:FlapperAngles}
	\end{figure}%
	\begin{table}[tb]
		\centering
		\caption{Experimental parameters of the flexible membrane wings in this study. The experiments are conducted in water with $\kindex{\nu}{\SI{20}{\celsius}} = \SI{1.00e-06}{\meter\squared\per\second}$.}
		\label{tab:wingparam}
		\begin{tabular}{lcc}
			\hline
			Parameters & & model wing \\ \hline
			Wing chord & $c$ & \SI{55}{\milli\meter} \\
			Wing span & $R$ &  \SI{150}{\milli\meter} \\
			Membrane thickness & $h$ &  \SIrange{0.3}{1.4}{\milli\meter} \\
			Wing stroke frequency & $f$ & \SIrange{0.125}{0.4}{\hertz} \\
			Angle of attack & $\hat{\alpha}$ & \SIrange{15}{75}{\degree} \\
			Peak-to-peak stroke amplitude & \kindex{\phi}{A} & \ang{90} \\
			Reduced frequency & $k$ & \num{0.42} \\
			Reynolds number & $Re$ & \numrange{2800}{9300} \\ \hline
		\end{tabular}
	\end{table}%
	The wing kinematics in hovering flight are described in terms of their stroke angle $\phi$, the angle of attack $\alpha$, and the flap or elevation angle $\theta$ (\cref{fig:FlapperAngles}a).
	The flap or elevation angle $\theta$ moves the wing up or down in stroke-normal direction (not displayed in~\cref{fig:FlapperAngles}).
	In this study, the elevation angle is not considered and stays constant at $\theta = \ang{0}$.
	The stroke plane coincides with the horizontal plane in the laboratory's frame of reference and the stroke angle $\phi$ defines the motion of the wing in the stroke plane.
	The stroke angle profile varies sinusoidally from $\phi = \ang{-45}$ to $\phi = \ang{45}$ and changes in frequency between $f = \SI{0.125}{\hertz}$ and $f = \SI{0.4}{\hertz}$ (\cref{tab:wingparam}).
	The angle of attack $\alpha$ indicates the wing's rotation relative to the stroke plane and is defined as the angle between the chord length and the horizontal stroke plane as indicated in \cref{fig:BillyBoy}b,c.
	In this study, the wing rotation is symmetric with respect to the stroke motion such that the angle of attack equals \ang{90} at the start and end of the stroke.
	The angle of attack follows a trapezoidal profile and reaches minimum values ranging from $\hat{\alpha} = \ang{15}$ to $\hat{\alpha} = \ang{75}$.
	Throughout this paper, we use $\hat{.}$ to denote amplitudes or minimum or maximum values of different quantities.
	The angle $\hat{\alpha}$ is kept constant during \SI{68}{\percent} of the total cycle duration $T=1/f$.
	During the remaining \SI{32}{\percent} of the cycle, the wing reverses orientation between the symmetric front- and back-strokes, this is often referred to as the flip duration~\cite{sane_control_2001, krishna_flowfield_2018}.\\
	Non-dimensional numbers characterising the flow around the hovering wing are the Reynolds number \textit{Re} and the reduced frequency $k$.
	The fluid-structure interactions of the compliant membranes are characterised by the Weber number \textit{We} and the aeroelastic number \textit{Ae} defined in equations \ref{eq:weber_number} and \ref{eq:aeroelastic_number}~\cite{song_aeromechanics_2008, waldman_camber_2017}.\\
	The Reynolds number $Re$ defines the ratio between inertial and viscous forces as a measure for the emergence of flow structures at different length and time scales.
	In hovering flight, the Reynolds number can be calculated as
	\begin{equation}
		Re=\frac{\overline{\U} c}{\nu} = \frac{2 \kindex{\phi}{A} f c \kindex{R}{2}}{\nu} \quad ,
		\label{eq:reynolds_number}
	\end{equation}%
	where $\nu$ denotes the kinematic viscosity of the fluid and $\overline{\U} = 2 \kindex{\phi}{A} f \kindex{R}{2}$ is the stroke-average wing velocity at the radius of second moment of area \kindex{R}{2}~\cite{sane_control_2001, sum_wu_scaling_2019, gehrke_phenomenology_2021}.
	The radius of second moment of area is the span-wise position where the sum of all forces apply and is $\kindex{R}{2} = \sqrt{\int_0^R(\kindex{R}{0}+r)^2 dr/R}$ for a rectangular wing planform~\cite{gehrke_phenomenology_2021}.
	The Reynolds number for all presented experiments in \cref{tab:wingparam} varies from \textit{Re} = \numrange{2800}{9300} which is a range where larger flying insects like the Hawkmoth (\emph{Manduca sexta}), small birds like the Rufous Hummingbird (\emph{Selasphorus rufus}), and several bat species like the Pallas's long-tongued bat (\emph{Glossophaga soricina}) take flight~\cite{shyy_recent_2010, muijres_leading-edge_2008}.\\
	The reduced frequency $k$ compares the spatial wavelength of the flow disturbance to the chord length $c$ and is a metric for the unsteadiness of the flow:
	\begin{equation}
		k=\frac{\pi c}{2 \kindex{\phi}{A} \kindex{R}{2}} \quad .
		\label{eq:reduced_frequency}
	\end{equation}%
	Here, $\kindex{\phi}{A} = \ang{90}$ is the peak-to-peak stroke amplitude.
	The reduced frequency of all our experiments is $k = \num{0.42}$ which is similar to many hovering flapping wing fliers in nature~\cite{shyy_recent_2010} and is considered to give raise to highly unsteady aerodynamics.
	\subsection{Flapping wing platform}
	The aerodynamic performances of the different membrane wings and the rigid reference case are evaluated experimentally with a robotic flapping wing device.
	The experiments are conducted in an octagonal tank with an outer diameter of \SI{0.75}{\metre} filled with water at a temperature of \SI{20}{\celsius}.
	The mechanism is controlled by two servo motors (Maxon motors, type RE35, \SI{90}{\watt}, \SI{100}{\newton\milli\metre} torque, Switzerland) which guide the stroke and pitch axis of the system.
	The motors are equipped with planetary gear-heads of $35:1$ and $19:1$ reduction for stroke and pitch respectively and are controlled using a motion controller (DMC-4040, Galil Motion Control, USA).
	The experimental flapping wing apparatus is designed to be highly repeatable and robust over a large number of experiments.
	The system allows for complex time-varying kinematics to be executed on both motors and initial tests on all frequencies $f$ and amplitudes $\hat{\alpha}$ show a maximum error of $< \ang{0.1}$ between the motor control signal and the motor response recorded by the encoder throughout the entire stroke.\\
	The aerodynamic loads on the wing are measured using a six-axis	IP68 force-torque transducer (Nano17, ATI Industrial Automation, USA) mounted at the wing root with a resolution of \SI{3.13}{\milli\newton} for the force and \SI{0.0156}{\newton\milli\metre} for torque measurements.
	The force signals from the load cell are recorded at a sampling frequency of \SI{1000}{\hertz} with a data acquisition module (NI-9220, National Instruments, USA).
	The instantaneous lift $L$, drag $D$ and pitch torque \kindex{T}{p} are directly retrieved from the load transducer.
	Here, the lift force $L$ is considered to be the component of the total force vector oriented upwards, perpendicular to the horizontal stroke plane.
	The drag $D$ is the force component in the stroke plane.
	The drag component is considered positive when it acts in the direction of the instantaneous velocity $U$ of the flow experienced by wing during its stroke motion (\cref{fig:BillyBoy}b and c).
	The aerodynamic power $P$ is determined as the sum of pitch power \kindex{P}{p} and stroke power \kindex{P}{s}.
	The power expended to rotate the wing around its pitch axis with an angular velocity $-\dot{\alpha}$ is determined by $\kindex{P}{p} = -\kindex{T}{p} \dot{\alpha}$.
	Analogously, the power required to rotate the wing around its stroke axis is found with the stroke torque \kindex{T}{s} and the stroke velocity $\dot{\phi}$ according to $\kindex{P}{s} = \kindex{T}{s} \dot{\phi}$.
	We cannot measure the stroke torque directly but instead determine it from the drag force $D$ across the span $\kindex{T}{s} = \int_R D(r) r \, dr$~\cite{sane_control_2001}.
	The stroke torque is calculated as $\kindex{T}{s} = D \kindex{R}{d}$ and the radial position $\kindex{R}{d} = \frac{3}{4}\frac{(\kindex{R}{0}+R)^4-\kindex{R}{0}^4}{(\kindex{R}{0}+R)^3-\kindex{R}{0}^3}$ where the sum of the drag force applies assuming a uniform drag coefficient distribution across the span~\cite{gehrke_phenomenology_2021}.\\
	Forces and torques are recorded over $16$ consecutive cycles.
	The first $5$ cycles are discarded to account for transient effects.
	The force and torque measurements are averaged over the remaining $11$ cycles to obtain phase-averaged temporal evolutions of the results within the flapping period $T$ and to determine the overall mean values of the aerodynamic coefficients.
	A minimum tip clearance of $3.5 c$ is found for a tip-to-tip stroke-amplitude of $\kindex{\phi}{A} = \ang{90}$ which has been shown sufficient to avoid wall effects in flapping wing experiments~\cite{manar_tip_2014, krishna_flowfield_2018}.
	Initial force measurements at different distances from the wall verify that for the selected stroke amplitude and frequencies no changes in the stroke-average forces are identified for wing tip to wall distances $>3 c$.
	Initial experiments collecting data for $64$ cycles demonstrate that there are some force variations due to recirculation of the tank but these only take effect after the first $16$ cycles that are considered here.
  The influence of the large-scale recirculation in the tank are below \SI{2}{\percent} compared to the mean force coefficients presented in this study (see \ref{sec:force_average_convergence} for more details).\\
	For high-frequency flapping wing flight in air, the wing inertial forces can be as strong as the aerodynamic pressure forces~\cite{gopalakrishnan_effect_2010}.
	We quantified the effect of the wing inertia on the total forces by conducting additional experiments in air (\ref{sec:wing_inertia}).
	Here, the aerodynamic forces become negligible and the inertial forces dominate the force measurements.
	The average lift and drag coefficients measured in air are largest for the high frequency cases but remain at least one order of magnitude lower than the results measured in water.
	For low flapping frequencies the dimensional inertial lift and drag are even lower and drop below the load cell resolution ($\SI{3.13}{\milli\newton}$).
 	The inertial forces are therefore deemed negligible in this study.
	\subsection{Stereo photogrammetry}
	In this work, we also perform stereo deformation measurements of the flexible wing platform to quantify the membrane shape and its influence on the aerodynamic forces.
	Deformation and load measurements are conducted simultaneously for selected experimental conditions to allow for direct comparison of deformation and aerodynamic force response.
	Two CCD cameras (\textit{pco.pixelfly usb}, ILA\_5150 GmbH/PCO AG, Germany) at a \ang{45} stereo angle and equipped with \SI{12}{\milli\meter} focal length lenses are used to perform marker tracking on the membrane wing throughout the entire flapping wing cycle (\cref{fig:FlapperAngles}b).
	With a camera resolution of $\SI{1392 x 1040}{\pixel}$, the deformation measurements have a spatial resolution between \SI{0.13}{\milli\metre\per\pixel}, when the wing is closest to the cameras, and \SI{0.23}{\milli\metre\per\pixel}, when the wing is farthest away from the cameras.
	We record stereo images over \num{20} cycles.
	The first \num{5} cycles are discarded to remove transient effects.
	The deformation measurements of the remaining \num{15} cycles are phase-averaged over one half-cycle, using the symmetry between front- and back-stroke of the flapping motion~\cite{gehrke_phenomenology_2021}.
	The images are recorded at an acquisition frequency of $\SI{6.17}{\hertz}$ uncorrelated to the flapping frequency $f$.
	In consecutive cycles the images correspond to different phase-times $t/T$ within one full cycle $T$.
	After sorting the frames by their relative phase-time we achieve an increased image acquisition rate of $\SI{186}{\hertz}$ per half-cycle or between \num{465} and \num{1488} times the flapping frequency $f$.\\
	The membrane is covered with black markers and the carbon fibre and structural parts with white markers to ensure high contrast (\cref{fig:FlapperAngles}a).
	The markers are tracked using the software tool \textit{XMALab}, an open-source software for marker-based X-ray reconstruction of moving morphologies~\cite{knorlein_validation_2016}.
	The stereo camera configuration is calibrated using a chequerboard recorded at different positions and rotation angles for both cameras.
	The calibration procedure determines all camera intrinsic parameters, calibration coefficients and the image distortion matrix.
	The markers are tracked at each time step in the undistorted stereo images and their position in three-dimensional space reconstructed using linear triangulation~\cite{mikhail_introduction_2001}.
	The orientation and position of the rigid parts of the wing platform are determined through rigid body transformation.
	Finally, we determine the membrane shape with a two-dimensional polynomial which fits best the membrane markers and the angle between the membrane and the rigid edges of the wing in a least-square sense.
	\section{Results}%
	The aerodynamic performance of the new adaptive membrane wing is evaluated in terms of the stroke-average lift coefficient $\kindex{\overline{C}}{L}$ and its hovering efficiency $\overline{\eta}$~\cite{gehrke_genetic_2018, gehrke_phenomenology_2021}:
	\begin{equation}
		\kindex{C}{L} = \frac{L}{\frac{1}{2}\rho R c \overline{U}^2} , \quad \kindex{C}{P} = \frac{P}{\frac{1}{2}\rho R c \overline{U}^3} , \quad \overline{\eta} = \frac{\kindex{\overline{C}}{L}}{\kindex{\overline{C}}{P}} \quad ,
		\label{eq:aerodynamic_coefficients}
	\end{equation}
	where $L$ and $P$ are the dimensional lift and power respectively.
	Here, $\overline{\rule{0ex}{1ex}.}$ denotes stroke-average quantities and $\overline{U} = 2 \phi f \kindex{R}{2}$ is the stroke-average wing velocity computed at the radius of the second moment of area $\kindex{R}{2}$.
	The average lift coefficient determines how much weight the flapping wing system can support or how fast it can climb in altitude.
	The hovering efficiency limits the total flight time of the vehicle.
	\subsection{Overview of the phase-average performance of the membrane and rigid wings}
	\begin{figure}
		\centering
		\includegraphics[]{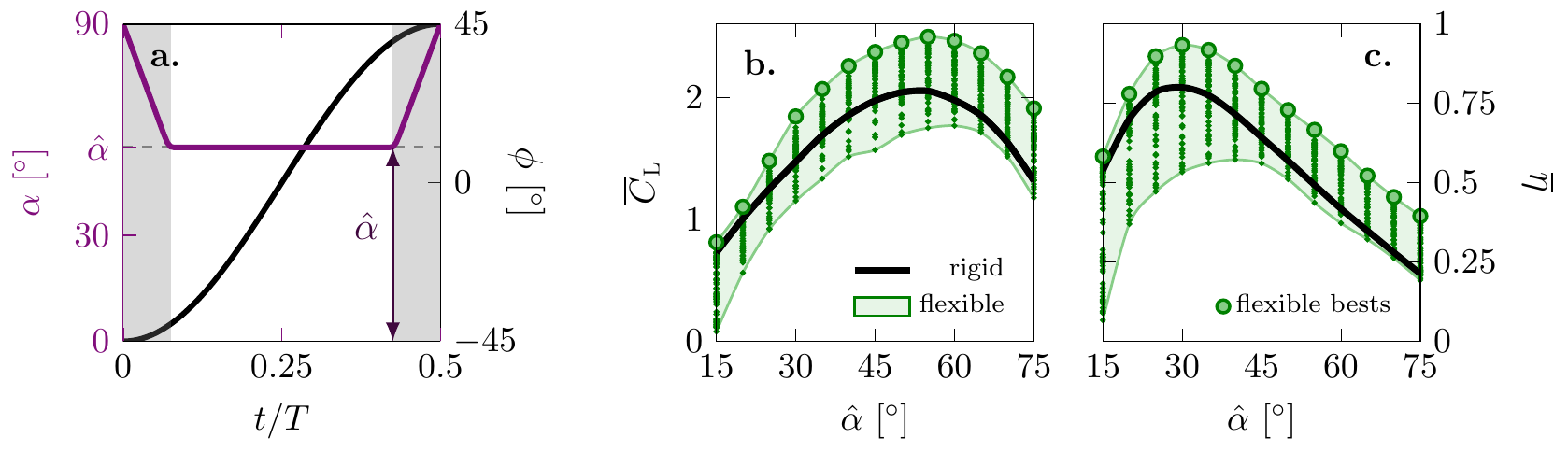}
		\caption{a.~Temporal variation of the angle of attack $\alpha$ and stroke angle $\phi$ over one half-cycle, b.~stroke-average lift coefficient as a function of the angle of attack, and c.~hovering efficiency as a function of the angle of attack. The shaded envelopes in b. and c. cover all conducted experiments.}
		\label{fig:forcesVSAlpha}
	\end{figure}%
	The flapping wing kinematics in all experiments are defined by the stroke and pitch angle profile presented in \cref{fig:forcesVSAlpha}a.
	Here, only the first half of the cycle is presented as the front- and back-strokes of the flapping wing motion are symmetric.
	At the beginning and end of each half-cycle the wing rotates to keep the leading edge ahead of the trailing edge.
	The duration of the wing rotation is indicated by the grey area in \cref{fig:forcesVSAlpha}a.
	During the majority of the cycle, the angle of attack remains at a constant value of $\hat{\alpha}$.
 	For brevity, we will refer to the constant minimum angle of attack during the stroke motion simply as the angle of attack.
	The system's base frequency $f$ and the angle of attack $\hat{\alpha}$ are varied over a wide range shown in \cref{tab:wingparam} to characterise the aerodynamic performance of the different membrane wings in comparison to rigid reference cases.\\
	The stroke-average lift coefficient \kindex{\overline{C}}{L} and hovering efficiency $\overline{\eta}$ for all tested flapping frequencies $f$, wing thicknesses $h$, and angles of attack $\hat{\alpha}$ are presented in \cref{fig:forcesVSAlpha}b and c.
	The black curve connects the results for the rigid reference case averaged over all stroke frequencies $f$.
	As the rigid wings do not deform, low variance in \kindex{\overline{C}}{L} and $\overline{\eta}$ is observed for each angle of attack $\hat{\alpha}$ compared to the membrane wings.
	The results of the best performing membrane wings are highlighted by the large green markers.
	The small markers indicate individual test cases.
	In general, the membrane wings reach maximum stroke-average lift \kindex{\overline{C}}{L} and efficiency $\overline{\eta}$ values in a range around the values attained by the rigid reference cases for the same angle attack $\hat{\alpha}$.
	By varying the thickness of the membranes, we cover a relevant parameter range in terms of membrane stiffness.
	Our set of tested membrane wings include membranes that perform better and some that perform worse than their rigid counterparts.
 	For the lowest angle of attack $\hat{\alpha}=\ang{15}$, all membrane wings tested at best match the performance of the rigid wing.
 	For the highest angle of attack $\hat{\alpha}=\ang{75}$, all membrane wings perform equally well or better than the rigid wing.\\
	The highest stroke-average lift is found at $\hat{\alpha} = \ang{55}$ with $\kindex{\overline{C}}{L,max} = \num{2.43}$ for the membrane wings and at $\hat{\alpha} = \ang{50}$ with $\kindex{\overline{C}}{L,max} = \num{2.06}$ for the rigid wing.
	The variance in aerodynamic performance of the membrane at each angle of attack $\hat{\alpha}$ in \cref{fig:forcesVSAlpha}a and b is due to the deformation of the membrane which depends on the ratio between the dynamic pressure on the wing as a function of the flapping frequency $f$ and the rigidity or bending stiffness of the membrane which varies with membrane thickness $h$.
	The aeroelastic number \textit{Ae} (\cref{eq:aeroelastic_number}) represents the ratio between membrane compliance and the dynamic pressure and is used to further characterise the influence of the fluid-membrane interaction on the aerodynamic performance of the membrane wings.
	Note that no additional pre-tension is applied to the membrane or the trailing edge sliders that could influence the effective stiffness of the system.
	\subsection{Lift enhancement through deformation}
	\begin{figure}
		\centering
		\includegraphics[]{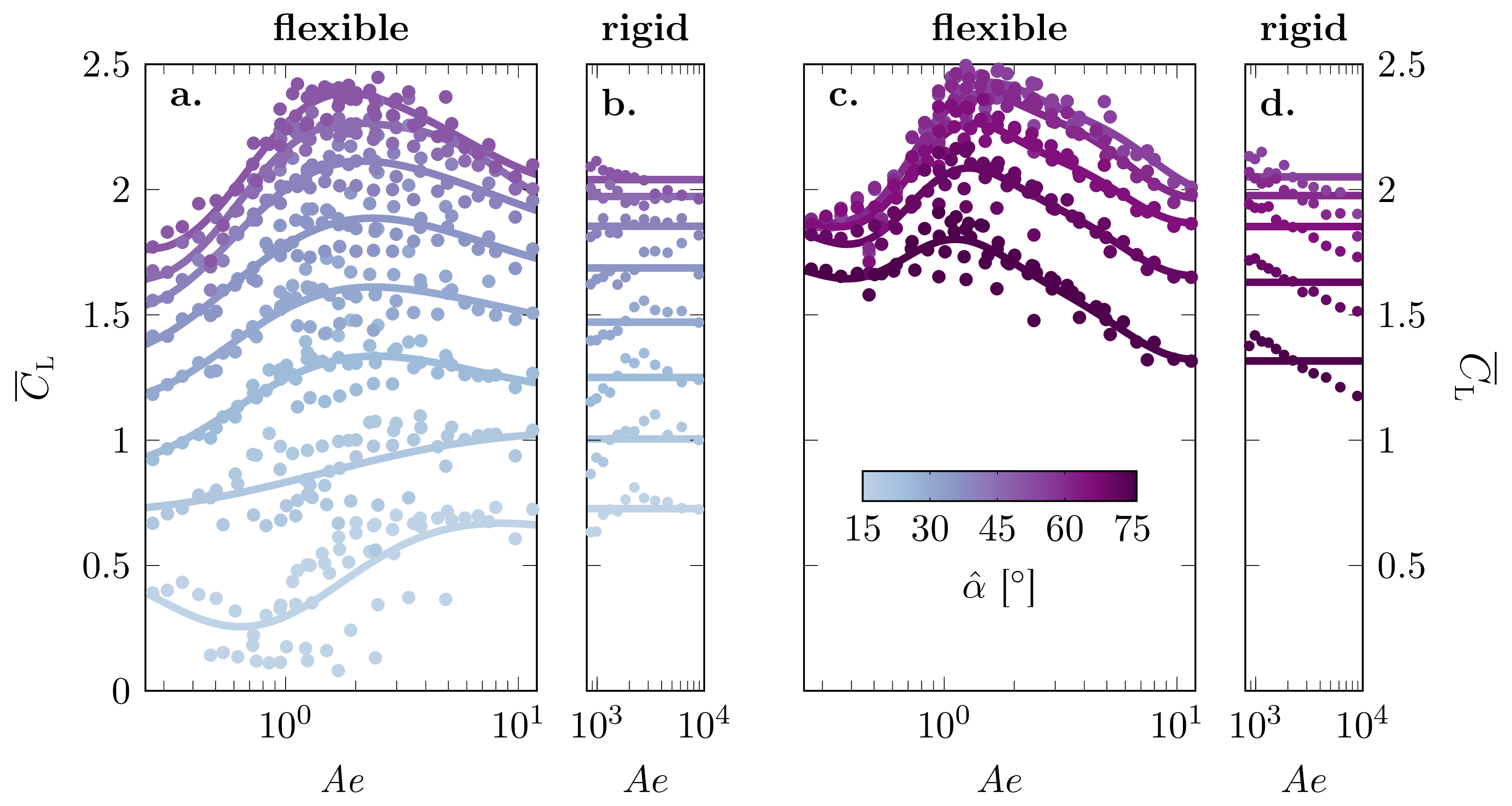}
		\caption{
			Stroke-average lift coefficient \kindex{\overline{C}}{L} for different angles of attack $\hat{\alpha}$ as a function of the aeroelastic number \textit{Ae}.
			Results for angles of attack ranging from $\hat{\alpha} = \ang{15}$ to $\ang{50}$ are presented in a and b.
 			Results for higher angles of attack ranging from $\hat{\alpha} = \ang{55}$ to $\ang{75}$ are presented in c and d.
			Each dot represents an individual experiment.
			The solid lines indicate the average variation of \kindex{\overline{C}}{L} as a function of \textit{Ae} for each angle of attack.
		}
		\label{fig:CLVSAe}
	\end{figure}%
	The lift produced in hovering flight is presented in \cref{fig:CLVSAe} as a function of the aeroelastic number \textit{Ae} (\cref{eq:aeroelastic_number}) for all conducted experiments.
	Overall, the stroke-average lift of the membrane wings increases with increasing angle of attack for $\hat{\alpha} \leq \ang{50}$ and decreases for $\hat{\alpha} \geq \ang{55}$.
	For visual clarity, the results are split in different panels in \cref{fig:CLVSAe}.
	The two panels to the left (\cref{fig:CLVSAe}a and b) include all experiments with angles of attack ranging from $\hat{\alpha} = \ang{15}$ to $\ang{50}$ for the flexible membrane wings (\textit{Ae} = \numrange{0.25}{12} in \cref{fig:CLVSAe}a) and the rigid wings (\textit{Ae} = \numrange{870}{8900} in \cref{fig:CLVSAe}b).
	The markers represent individual experiments.
	The solid lines indicate the average variation of \kindex{\overline{C}}{L} as a function of \textit{Ae} and are obtained by training a Gaussian process regression model at each angle of attack $\hat{\alpha}$ and predicting the response over the \textit{Ae} range of the experiments.
	The stroke-average lift coefficient \kindex{\overline{C}}{L} for $\ang{20}\leq\hat{\alpha}\leq\ang{50}$ increases with increasing \textit{Ae} until it reaches a maximum between $\textit{Ae} = \num{1}$ and $\textit{Ae} = \num{5}$.
	The maximum stroke-average lift coefficient \kindex{\overline{C}}{L} is reached at slightly lower values of the aeroelastic number at higher angles of attack.
	For $\textit{Ae}>5$ and $\hat{\alpha}>\ang{20}$, the stroke-average lift decreases with increasing \textit{Ae} and asymptotically converges to the average lift produced by the rigid wings (\cref{fig:CLVSAe}b).

	At lower angles of attack $\hat{\alpha} \leq \ang{20}$, the maximum stroke-average lift values are obtained for the stiff wings and the flexible wings with the highest values of \textit{Ae}.
	For the lowest angle of attack $\hat{\alpha}=\ang{15}$, the stroke-average lift initially decreases with increasing \textit{Ae} for $\textit{Ae}<1$ and we notice larger variations among individual experiments.
	Observations of the experiments showed that the cambering of the membrane wings started later into the cycle compared to higher angles of attack and that the amount of camber showed more variability between consecutive strokes.
	Additional experiments at the lower angles of attacks are required to determine if this effect is caused by unsteady aeroelastic effects or limitations of the experimental system.
	For higher angles of attack $\hat{\alpha} > \ang{50}$, the overall stroke-average lift decreases with increasing angles of attack (\cref{fig:CLVSAe}c and d).
	At these high angles of attack, the flexible wings attain values of the stroke-average lift above those attained by their solid counterparts for all values of the aeroelastic number.
	The aeroelastic number at which the overall maximum stroke-average lift coefficient is reached remains around $\textit{Ae}\approx 1$ and continues to decreases with increasing angles of attack.
	\\
	\begin{figure}
		\centering
		\includegraphics[]{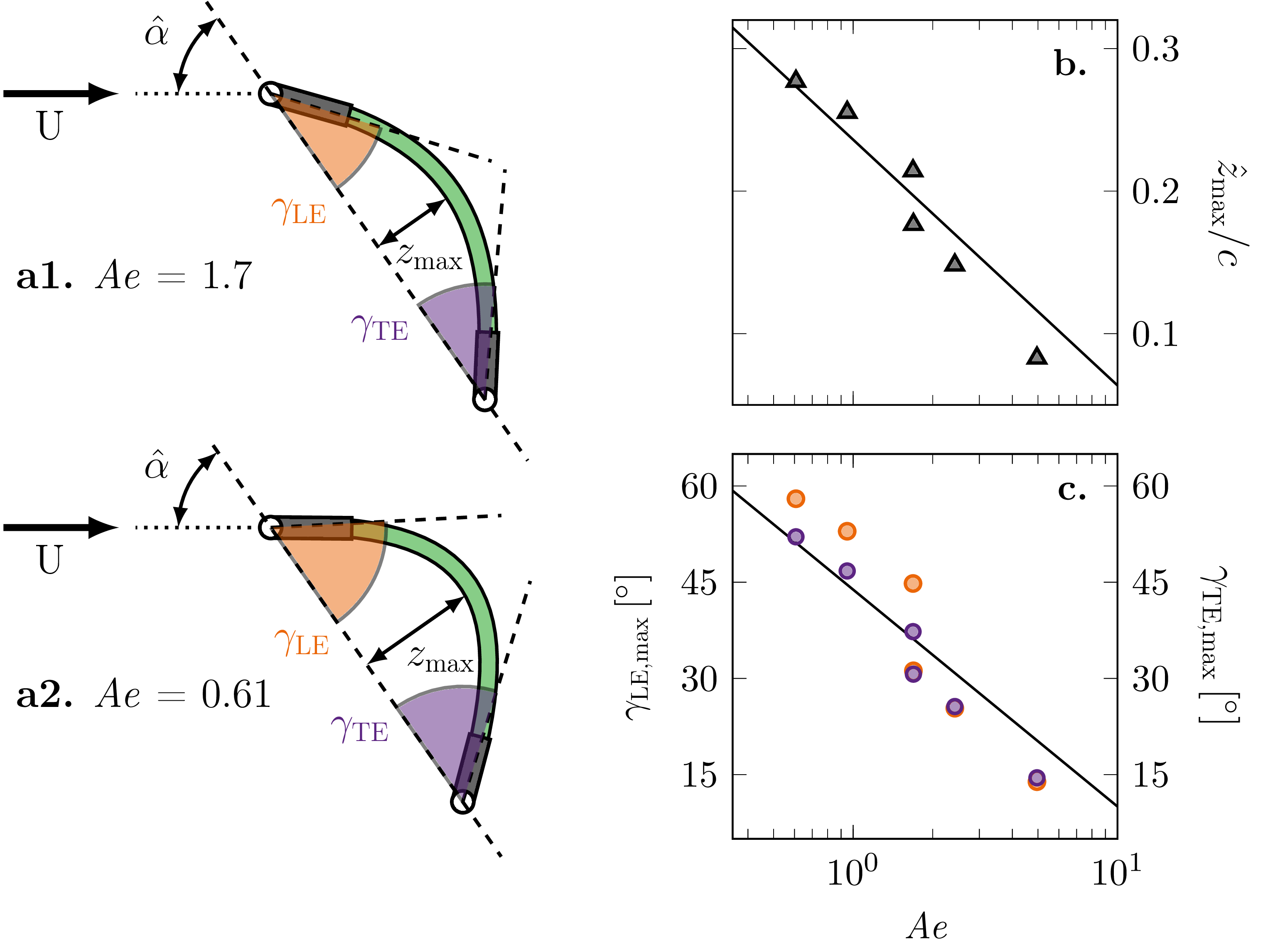}
		\caption{a.~Schematic of the wing deformation indicating the definition of the wing camber $\hat{z}$, the leading and trailing edge rotation angles \kindex{\gamma}{LE} and \kindex{\gamma}{TE} for $\textit{Ae}=1.7$ (a1) and $\textit{Ae}=0.61$ (a2).
		b.~Stroke-maximum camber \kindex{\hat{z}}{max} as a function of the aeroelastic number for $\hat{\alpha} = \ang{55}$.
		c.~Stroke-maximum leading and trailing edge rotation angles \kindex{\gamma}{LE,max} and \kindex{\gamma}{TE,max} as a function of the aeroelastic number for $\hat{\alpha} = \ang{55}$.}
		\label{fig:camberGammaVSAe}
	\end{figure}%
  To explain the differences observed among the stroke-average lift and efficiency measured of the flapping membrane and rigid wings, we have extracted the membrane shape deformation for selected wings.
	The membrane wings passively deform when flapping.
	The main characteristics of the passive deformation are the passive cambering of the membrane, the rotation of the leading edge and the rotation of the trailing edge (\cref{fig:camberGammaVSAe}a).
	The stroke-maximum camber \kindex{\hat{z}}{max} and the stroke-maximum rotation angles of the leading and trailing edges with respect to the chord, denoted by \kindex{\gamma}{LE,max} and \kindex{\gamma}{TE,max}, respectively, are presented as a function of the aeroelastic number in \cref{fig:camberGammaVSAe}b-c for $\hat{\alpha}=\ang{55}$.
 	Positive angles of $\gamma$ are associated with a positive camber $\hat{z}$ following the convention indicated in \cref{fig:camberGammaVSAe}a.
	The stroke-maximum values reveal a clear variation with the aeroelastic number which relates the stiffness of the membrane and the aerodynamic pressure.
	At low values of the aeroelastic number, a relatively smaller amount of aerodynamic pressure is required to deform the wing.
	This results in large values of the stroke-maximum camber and the stroke-maximum leading and trailing edge rotation angles.
	With increasing values of the aeroelastic number, the effective stiffness of the wing increases and the maximum camber and maximum leading and trailing edge rotation angles decrease towards zero for values of $\textit{Ae}\gg 1$.\\
	The decrease of the stroke-maximum camber \kindex{\hat{z}}{max} with increasing aeroelastic number follows an exponential decay as evidenced by the linear evolution in the semi-log space.
	Exponential fits in form of straight lines are included in the semi-logarithmic plots as a reference.
	The decrease of the leading and trailing edge rotation angles also appear close to linear in the semi-log space at first glance, but a closer look reveals a more complicated variation.
	The leading and trailing edge rotation angles are identical for $\textit{Ae} \geq 1.68$, indicating symmetric cambering at higher values of the aeroelastic number.
	For lower values of the aeroelastic number and thus reduced effective membrane stiffness, the fluid-membrane interaction leads to asymmetric bending of the membrane and a shift of the maximum camber position towards the leading edge where the rotation angle is largest.
	The asymmetric bending is the result of a more complex fluid structure interaction between the aerodynamic pressure due to the stroke motion and a prominent leading edge vortex.
	The change between symmetric and asymmetric camber lines occurs between $\textit{Ae}=1$ and $\textit{Ae}=2$ which corresponds to the range of aeroelastic numbers where we observe a maximum in the stroke-average lift coefficient for this angle of attack ($\hat{\alpha}=\ang{55}$).
	The lift-optimal stroke-maximum camber in this case is $\kindex{\hat{z}}{max} \approx \num{0.2} c$.
	An increase in the maximum camber $\kindex{\hat{z}}{max}/c$ beyond \num{0.2} results in a decrease of the stroke-average lift.
 	To better understand why, we will now look at the orientation of the leading and trailing edges with respect to the stroke plane.

	\begin{figure}
		\centering
		\includegraphics[]{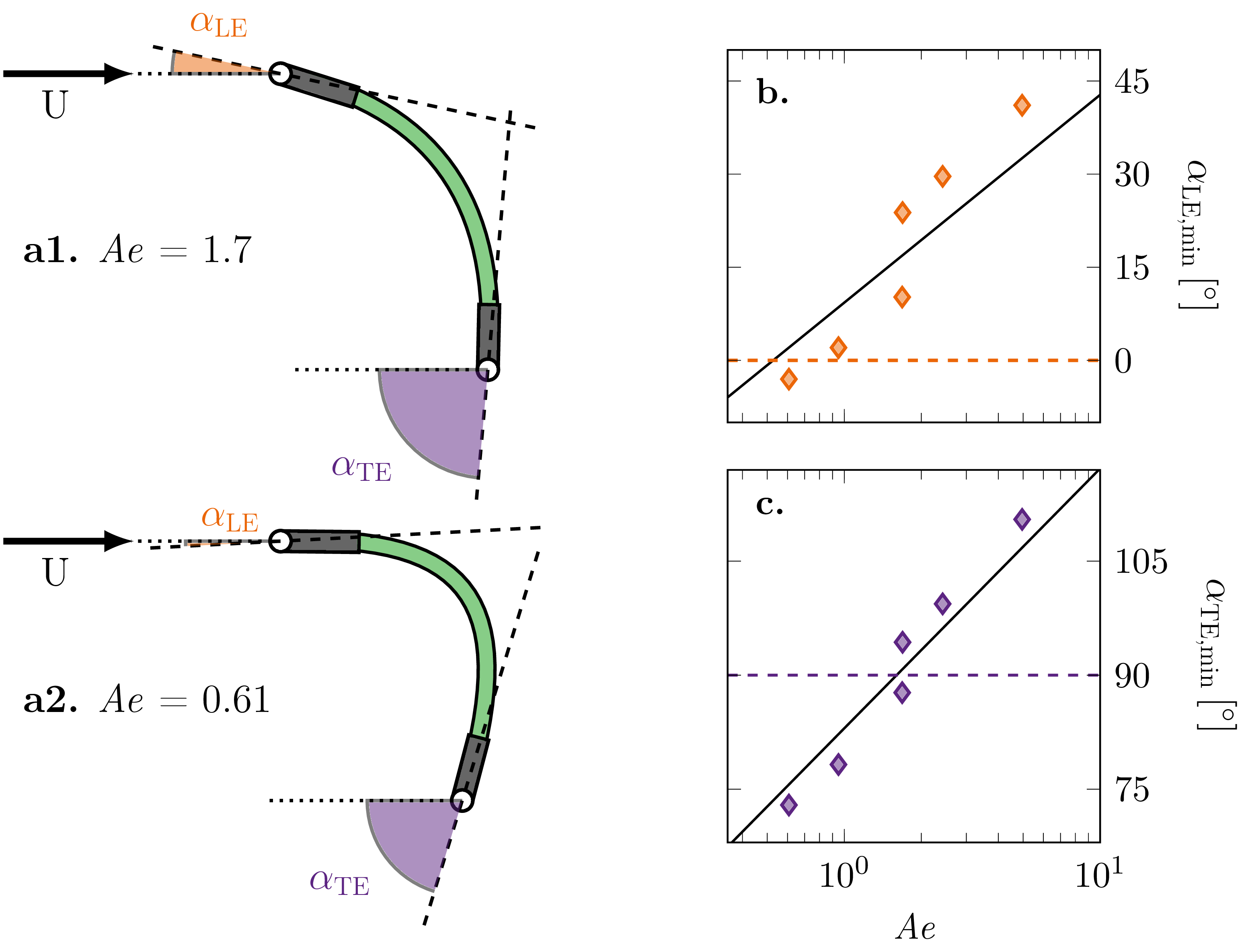}
		\caption{a.~Schematic of the wing deformation indicating the definition of the leading and trailing edge angles with respect to the stroke plane, \kindex{\alpha}{LE} and \kindex{\alpha}{TE} for $\textit{Ae}=1.7$ (a1) and $\textit{Ae}=0.61$ (a2). b.~Stroke-minimum leading edge angle \kindex{\alpha}{LE,min} as a function of the aeroelastic number for $\hat{\alpha} = \ang{55}$. c.~Stroke-minimum trailing edge angle \kindex{\alpha}{TE,min} as a function of the aeroelastic number for $\hat{\alpha} = \ang{55}$.}
		\label{fig:anglesVSAe}
	\end{figure}%
	The passive cambering of the membrane wings and the rotation of the leading and trailing edges lead to a decrease in the leading edge angle of attack with respect to the stroke plane and a decrease in the angle between the trailing edge and the stroke plane as defined in \cref{fig:anglesVSAe}a.
	The stroke-minimum leading and trailing edge angles \kindex{\alpha}{LE,min} and \kindex{\alpha}{TE,max} are presented as a function of the aeroelastic number in \cref{fig:anglesVSAe}b,c for $\hat{\alpha}=\ang{55}$.
	As expected based on the measured decay of the maximum camber, the minimum leading and trailing edge angles both increase with increasing aeroelastic number in an approximately exponential manner.
	For the higher values of the aeroelastic number, where the deformation is symmetric, the angles lie above the fitted approximations.
	For the lower values of the aeroelastic number, where the deformation is no longer symmetric, we find angles below the fitted approximation.\\
	The maximum stroke-average lift coefficient and the onset of asymmetric membrane deformation occur at the same aeroelastic number of $\textit{Ae}\approx \num{1.7}$.
	The measured membrane shape at these lift-optimal aeroelastic conditions is shown in \cref{fig:anglesVSAe}a1.
	The rotation of the leading edge has reduced the effective angle of attack at the leading edge to a more moderate angle and the trailing edge is oriented vertically with respect to the stroke plane.
	The reduced leading edge angle of attack allows the flow to accelerate smoothly around the leading edge and the trailing edge orientation directs the flow straight downwards in the near wake increasing the vertically upward force on the wing itself.\\
	By decreasing the aeroelastic number \textit{Ae} for the same angle of attack $\hat{\alpha}$, the deformation of the membrane increases but the stroke-average lift decreases.
	This is attributed to the trailing edge angle decreasing below $\kindex{\alpha}{TE,min} < \ang{90}$ (dashed line in \cref{fig:anglesVSAe}c) and the leading edge angle decreases to values below the static stall angle of a flat plate which delays flow separation and the formation of a strong leading edge vortex.
	The stroke-maximum membrane deformation at $\textit{Ae} = 0.61$ is presented in \cref{fig:anglesVSAe}a2.
	At this low value of the aeroelastic number, the chord-normalised camber becomes as high as \SI{30}{\percent} and the leading edge angle even becomes negative which causes the lift coefficient to drop below the values of their stiff counterparts.
	The stroke-average lift of our flapping wings is increased when they are deformed up to a point when they over-camber.
 	Here, we consider a membrane to be over-cambered when its minimum leading edge angle $\kindex{\alpha}{LE,min}$ becomes negative or its minimum trailing edge angle $\kindex{\alpha}{TE,min}$ rotates below \ang{90} as marked by the dashed lines in \cref{fig:anglesVSAe}b,c.
 	The performance of over-cambering wings rapidly falls with decreasing aeroelastic number.\\
	\subsection{Temporal evolution the wing deformation and aerodynamic performance}
	\begin{figure}[tb]
		\centering
		\includegraphics[]{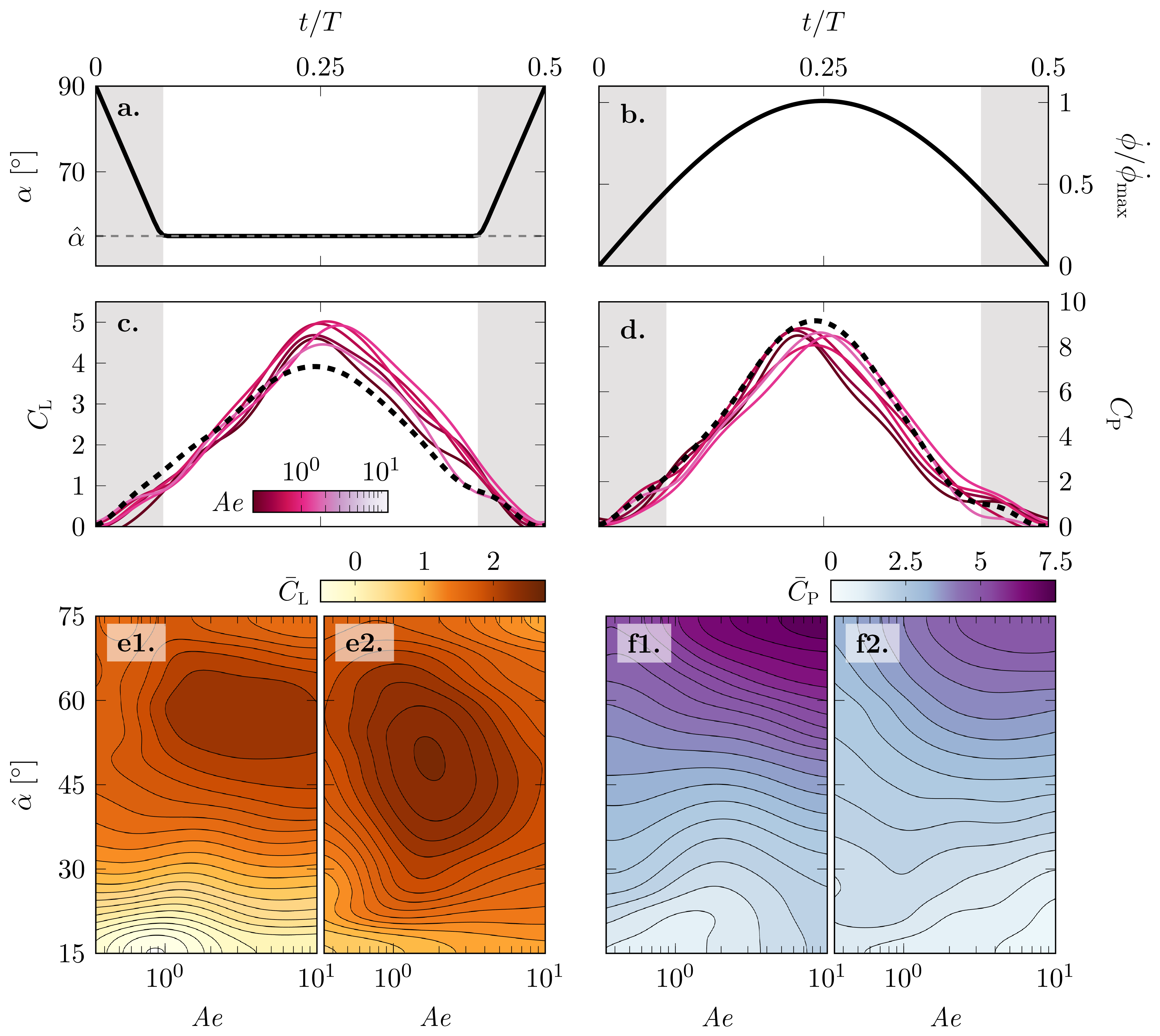}
		\caption{Temporal evolution of the flapping kinematics in terms of the angle of attack ($\alpha$) (a.) and stroke rate ($\dot{\phi}$) (b.), and the phase-averaged lift (c.) and power coefficient (d.) for $\hat{\alpha} = \ang{55}$.
 		As the motion is symmetric, only one half cycle $0\leq t/T\leq 0.5$ is shown.
		The dashed black lines represents the rigid reference case with no deformation.
		The grey shaded areas indicate the duration of the wing rotation between strokes.
		Contour plots showing the averaged lift/power coefficient generated during the first half of the stroke (e1./f1.) and during the second half of the stroke (e2./f2.) for all variations of the angle of attack and aeroelastic number tested.}
		\label{fig:forceVStime}
	\end{figure}%
	We further explore the relationship between the aeroelastic properties and their contribution to the force production on the wing by analysing the temporal evolution of the lift and power coefficients for the highest lift producing angle of attack $\hat{\alpha} = \ang{55}$ in \cref{fig:forceVStime}c,d.
	The colours indicate the range of aeroelastic numbers from \textit{Ae} = \numrange{0.61}{4.96}.
	Here, all quantities are phase-averaged over one half-cycle ($t/T$ = \numrange{0}{0.5}) exploiting the symmetry between the front- and back-stroke of the hovering cycle~\cite{gehrke_phenomenology_2021}.
	The results of the rigid wing are presented by the black dashed line.
	The phase-averaged temporal evolutions of the lift and power were filtered using a 5th-order Butterworth filter with no phase delay at a cut-off frequency $12$ times higher than the flapping frequency $f$.\\
	The overall temporal evolution of the lift and the power coefficient for all wings is dominated by the evolution of the stroke velocity $\dot{\phi}$ (\cref{fig:forceVStime}b).
	Lift and power increase or decrease with increasing or decreasing stroke velocity and reach a maximum value around $t/T=0.25$.
	Interesting differences are observed when the aeroelastic number varies.
	The main differences in the lift coefficient evolution occur towards the end of the first half and during the second half of the stroke ($t/T>0.2$) (\cref{fig:forceVStime}c).
	The flexible wings all reach a higher maximum lift coefficient than their rigid counterpart.
	The highest value of \kindex{C}{L} is measured for an aeroelastic number of $\textit{Ae} = \num{1.64}$.
	The higher lift values for the flexible wings are maintained during the second half of the stroke.
	The competition for the top rankings among different aeroelastic numbers is decided in the second half of the stroke.
	This holds true for all angles of attack as evidenced by the contour plots of the average lift coefficient across the first and second half of the stroke (\cref{fig:forceVStime}e1,e2).
	The average lift coefficient is consistently lower during the first half of the stroke than during the second half for all values of the aeroelastic number and all angles of attack.
	The average lift coefficient during the first half does not vary substantially with the aeroelastic number for $\hat{\alpha}<\ang{45}$.
	For $\ang{45}<\hat{\alpha}<\ang{70}$, the average lift increases with increasing aeroelastic number for $\textit{Ae}<1$ and reaches a plateau for $\textit{Ae}>1$.
	The average lift coefficient during the second half shows a clear global maximum for $\hat{\alpha}\approx\ang{50}$ and $\textit{Ae}\approx\num{1.7}$.
	This corresponds to conditions that yield the overall best performance.\\
	The analysis of the power coefficient \kindex{C}{P} leads to the same conclusion that the performance difference by the deformable wings is made in the second half of the stroke (\cref{fig:forceVStime}d,f).
	All power curves follow a similar increase in power with increase in the stroke velocity in the first part of the stroke (\cref{fig:forceVStime}d).
	A maximum in \kindex{C}{P} is reached slightly before mid-stroke.
	The highest peak in power belongs to the rigid wing.
	In the second half of the cycle, the aerodynamic power curves associated with the deformable wings remain below the rigid wing curve for most of the second part of the stroke.
	The least amount of power is required by the lowest value of the aeroelastic number which is the most deformable wing.
	The least deformable wing with the highest value of \textit{Ae} requires approximately the same amount of power as the rigid wing.
	For all tested angles of attack $\hat{\alpha}$ and aeroelastic numbers, the first half of the stroke requires more power than the second (\cref{fig:forceVStime}f1,f2).
	The strongest dependence on the aeroelastic number is observed for angles $\hat{\alpha} > \ang{45}$ in the second half of the stroke (\cref{fig:forceVStime}f2).
 	Here, the power is drops significantly with decreasing values of \textit{Ae}.
	At low \textit{Ae}, the stronger cambering of the membrane leads to a decrease in the frontal area which leads to reduced drag and power coefficients.\\
	\begin{figure}
		\centering
		\includegraphics[]{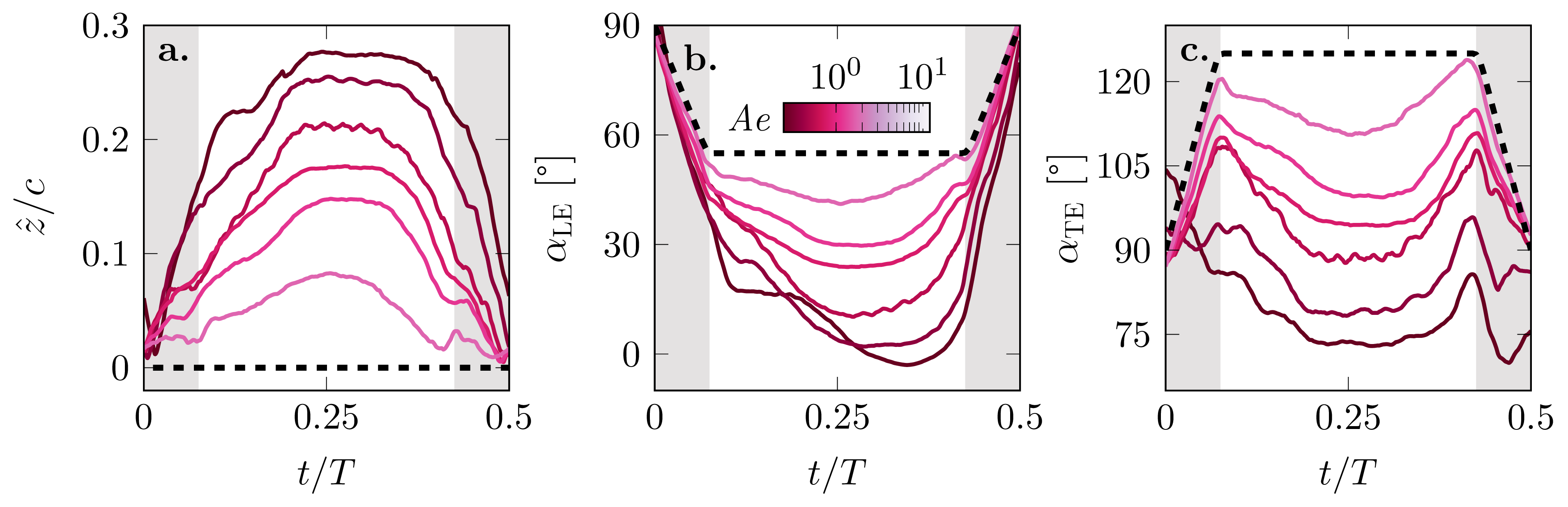}
		\caption{Temporal evolution of the wing deformation in terms of the wing camber ($\hat{z}$) and the leading and trailing edge angles with respect to the stroke plane ($\kindex{\alpha}{LE}$ and $\kindex{\alpha}{TE}$) for $\hat{\alpha} = \ang{55}$.
 		As the motion is symmetric, only one half cycle $0\leq t/T\leq 0.5$ is shown.
		The dashed black lines represents the rigid reference case with no deformation.
		The grey shaded areas indicate the duration of the wing rotation between strokes.}
		\label{fig:defVStime}
	\end{figure}%
	To explain the higher aerodynamic performance at moderate \textit{Ae} and the variations during the second half of the stroke, we analyse the temporal evolution of the passive membrane deformation characterised by the camber $\hat{z}$ and the leading and trailing edge angles \kindex{\alpha}{LE} and \kindex{\alpha}{TE} in \cref{fig:defVStime}.
	The data is presented for the lift-optimal angle of attack $\hat{\alpha} = \ang{55}$ over one stroke or half the flapping cycle.
  The results of the rigid wing are presented by the black dashed line.
	The line colour of the solid lines indicates the value of the aeroelastic number.\\
	The temporal evolutions of the membrane camber $\hat{z}$ for the deformable wings roughly follow a trapezoidal shape in response to the trapezoidal variation in their geometric angle of attack $\alpha$  (\cref{fig:defVStime}a).
	The camber for the lowest values of \textit{Ae} reaches the highest camber of $\hat{z}/c = \num{0.28}$ just before mid-stroke and keeps a steady camber over a larger portion of the second half of the stroke.
 	We refer to the time interval during which the camber remains at its maximum value as the camber plateau time.
	The time at which the maximum camber is reached does not vary significantly with effective membrane stiffness expressed by the aeroelastic number.
	The deformation seems to continue as long as the stroke velocity increases.
 	Yet, the rate of change of the camber during the first part of the stroke does not vary consistently with the stroke velocity and the response strongly depends on the value of \textit{Ae}.
	For the lowest values of \textit{Ae}, we measure an immediate increase of the camber already during the rotation phase.
	For the higher values of \textit{Ae}, the increase in camber becomes more prominent after the wing rotation has ended.
	This suggests that the rotational acceleration affects the initial response of the membrane, but we believe that it does not affect the maximum camber which was shown to scale well with the aeroelastic number (\cref{fig:camberGammaVSAe}a).
	The maximum camber decreases with increasing values of \textit{Ae} and this holds try for any value of the camber $\hat{z}$ at all times during the stroke.
 	The camber plateau time also reduces with increasing values of \textit{Ae}.
	At the highest presented aeroelastic number, the membrane cambers up to $\hat{z}/c = \num{0.08}$ and immediately reduces again during the second part of the stroke.
 	A plateau of constant camber is no longer observed.
	The persistence of the higher camber during the second part of the stroke at lower aeroelastic numbers leads to higher lift and a reduction of the power during that time (\cref{fig:forceVStime}).\\
	The temporal evolution of the phase-average leading and trailing edge angles with respect to the stroke plane are presented in \cref{fig:defVStime}b,c.
	When the membrane cambers more at lower \textit{Ae}, the leading and trailing edges angles drop below the corresponding reference values imposed on the rigid wing.
	All leading edge angles of attack \kindex{\alpha}{LE} in \cref{fig:defVStime}b reach a minimum angle during the wing's camber plateau time \cref{fig:defVStime}a.
	With increasing deformation at lower \textit{Ae}, the leading edge angle of attack minima decrease and are attained later in the second half of the stroke.
	At the lowest \textit{Ae}, the leading edge angle of attack becomes negative $\kindex{\alpha}{LE} = \ang{-3}$ at $t/T = 0.35$ in \cref{fig:defVStime}b.
	The excessive rotation of the leading edge due to over-cambering is accompanied by a decrease in lift coefficient \kindex{C}{L} but also by reduction in power \kindex{C}{P} (\cref{fig:forceVStime}c,d).\\
	The trailing edge angle \kindex{\alpha}{TE} evolves differently than the leading edge angle (\cref{fig:defVStime}c).
	The trailing edge orientation appears symmetric around mid-stroke and all curves of \kindex{\alpha}{TE} follow an M-shape evolution.
	Two peaks emerge at the end and the start of the wing rotations.
	After the first peak, the trailing edge angle decreases and the trailing edge rotates in the direction of the motion.
	For moderate to high values of \textit{Ae}, the angle remains constant for a part of the stroke and the duration of this constant angle phase increases again with decreasing aeroelastic number, similar to the behaviour of the camber plateau time.
	At lift-optimal aeroelastic conditions, the trailing edge orientation is perpendicular to the flow with $\kindex{\alpha}{TE} = \ang{90}$.
	Further increase in the camber $\hat{z}$, by lowering \textit{Ae}, causes the trailing edge to rotate into the flow at angles down to $\kindex{\alpha}{TE} = \ang{70}$.
	The trailing edge rotation $\kindex{\alpha}{TE} < \ang{90}$ indicates over-cambering and coincides with a decrease in lift production by the membrane wings.\\
	\subsection{Global optima of lift and efficiency}
	\begin{figure}
		\centering
		\includegraphics[]{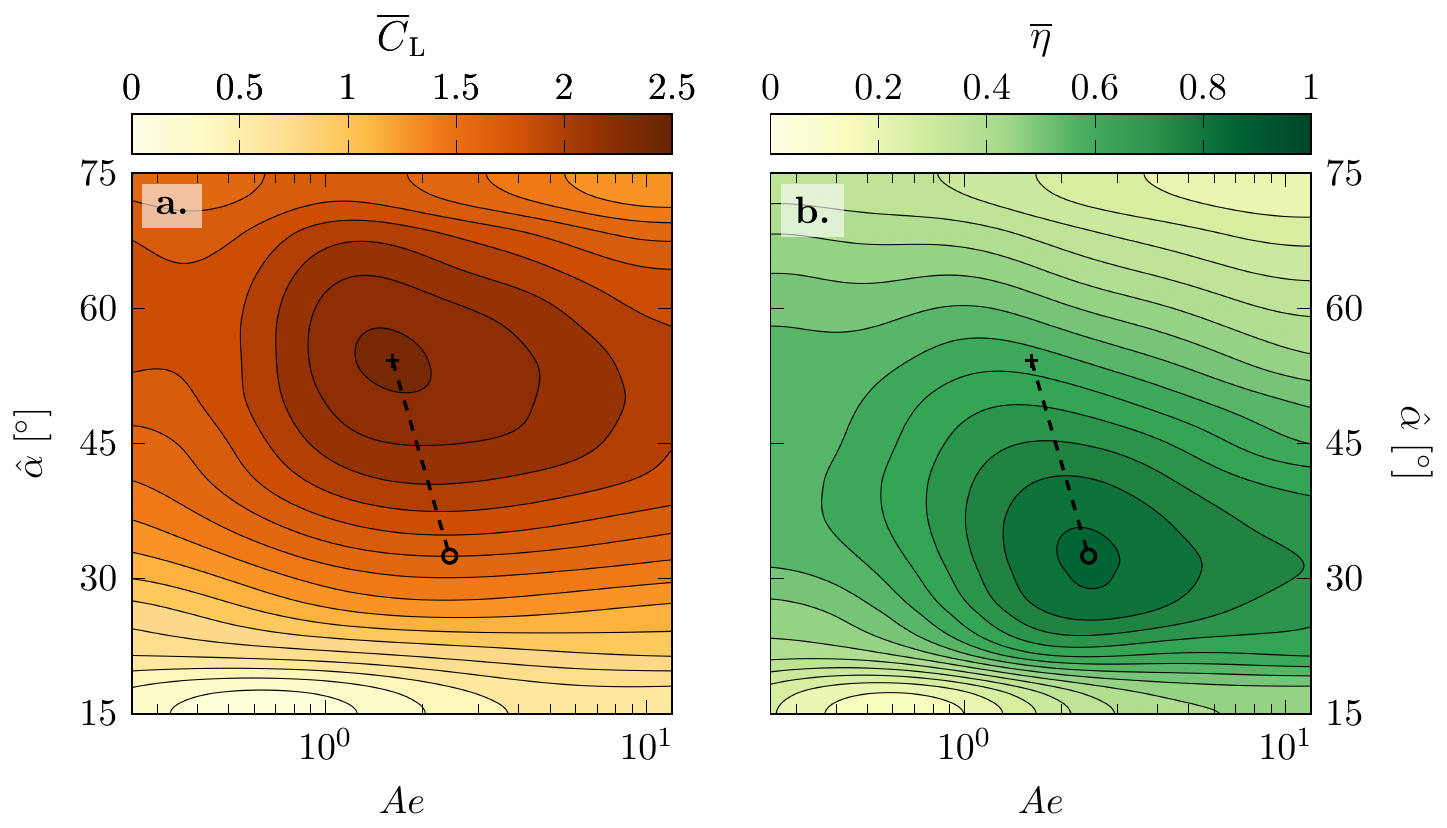}
		\caption{a.~Lift coefficient over aeroelastic number and angle of attack, and b.~hovering efficiency over aeroelastic number and angle of attack.}
		\label{fig:ClEtaVsAlphaAe}
	\end{figure}%
	In the previous sections, we analysed the relationship between lift coefficient \kindex{C}{L}, aeroelastic number \textit{Ae} for the lift-optimal angle of attack $\hat{\alpha}=\ang{55}$.
	In \cref{fig:ClEtaVsAlphaAe}, we summarise the results of the stroke-average lift coefficient \kindex{\overline{C}}{L} and hovering efficiency $\overline{\eta}$ (\cref{eq:aerodynamic_coefficients}) for the entire experimental parameter range covered in this study.
 	The input parameter space is spanned by the angle of attack $\hat{\alpha}$ and the aeroelastic number \textit{Ae}.
	The two markers in the contour plots indicate the optimal-lift ($\boldsymbol{\times}$) and optimal-efficiency ($\boldsymbol{\circ}$) points which lie at different locations.\\
	The maximum lift $\kindex{\overline{C}}{L,max} = \num{2.43}$ is found at $\textit{Ae} = \num{1.68}$ and $\hat{\alpha} = \ang{55}$ (\cref{fig:ClEtaVsAlphaAe}a).
	The most efficient hovering $\overline{\eta} = \num{0.858}$ occurs at a higher aeroelastic number $\textit{Ae} = \num{2.44}$ and a lower angle of attack $\hat{\alpha} = \ang{33}$ than the lift optimum (\cref{fig:ClEtaVsAlphaAe}b).
	In literature, we found an experimental study using compliant flat plate wings with a stiff leading edge that reports an enhancement of the aerodynamic performance of these flexible flapping wings in hover compared to rigid wings for a range of effective stiffness from $\kindex{\Pi}{1} = 0.5$ to $\kindex{\Pi}{1}=10$~\cite{fu_effects_2018}.
	The maximum lift production in \cite{fu_effects_2018} was found at $\kindex{\Pi}{1} = 3.5$.
	We find maximum lift production at a lower value of $\kindex{\Pi}{1} = 1.17$ using $\textit{Ae} = \kindex{\Pi}{1}^3$.
	The highest efficiency for our wings occurs at $\kindex{\Pi}{1} = 1.34$.
	Both optima still fall within the range of effective stiffness values that are reported to enhance aerodynamic performance by \cite{fu_effects_2018}.
	Compared to different insect species, the optimum chord-wise effective stiffness in our study matches with the tarantula-hawk wasp (\textit{Pepsis grossa}) which has similar aspect ratio wings than the membrane wings in our study~\cite{combes_flexural_2003, fu_effects_2018}.
	A recent study on membrane wings for energy harvesting applications discovered highest energy extraction at their lowest tested aeroelastic number $\textit{Ae} = 5$ ($\kindex{\Pi}{1} = 1.71$)~\cite{mathai_fluidstructure_2022}.\\
	%
	Rigid flapping wings at lower angles of attack promote the stable growth of a leading edge vortex and a lift favouring orientation of the normal force vector leading to more efficient hovering flight~\cite{gehrke_phenomenology_2021}.
	A shift of the most efficient hovering conditions to larger value of \textit{Ae} indicates a higher relative stiffness of the membrane and lower camber.
	The same leading edge deflection of up to $\kindex{\gamma}{LE,max} = \ang{45}$ observed at $\hat{\alpha} = \ang{55}$ and lift-optimal aeroelastic number $\textit{Ae} = \num{1.68}$ would lead to over-cambering and negative leading edge angles of attack \kindex{\alpha}{LE} at a geometric angle of attack $\hat{\alpha} = \ang{33}$ of most efficient hovering.\\
	\begin{figure}
		\centering
		\includegraphics[]{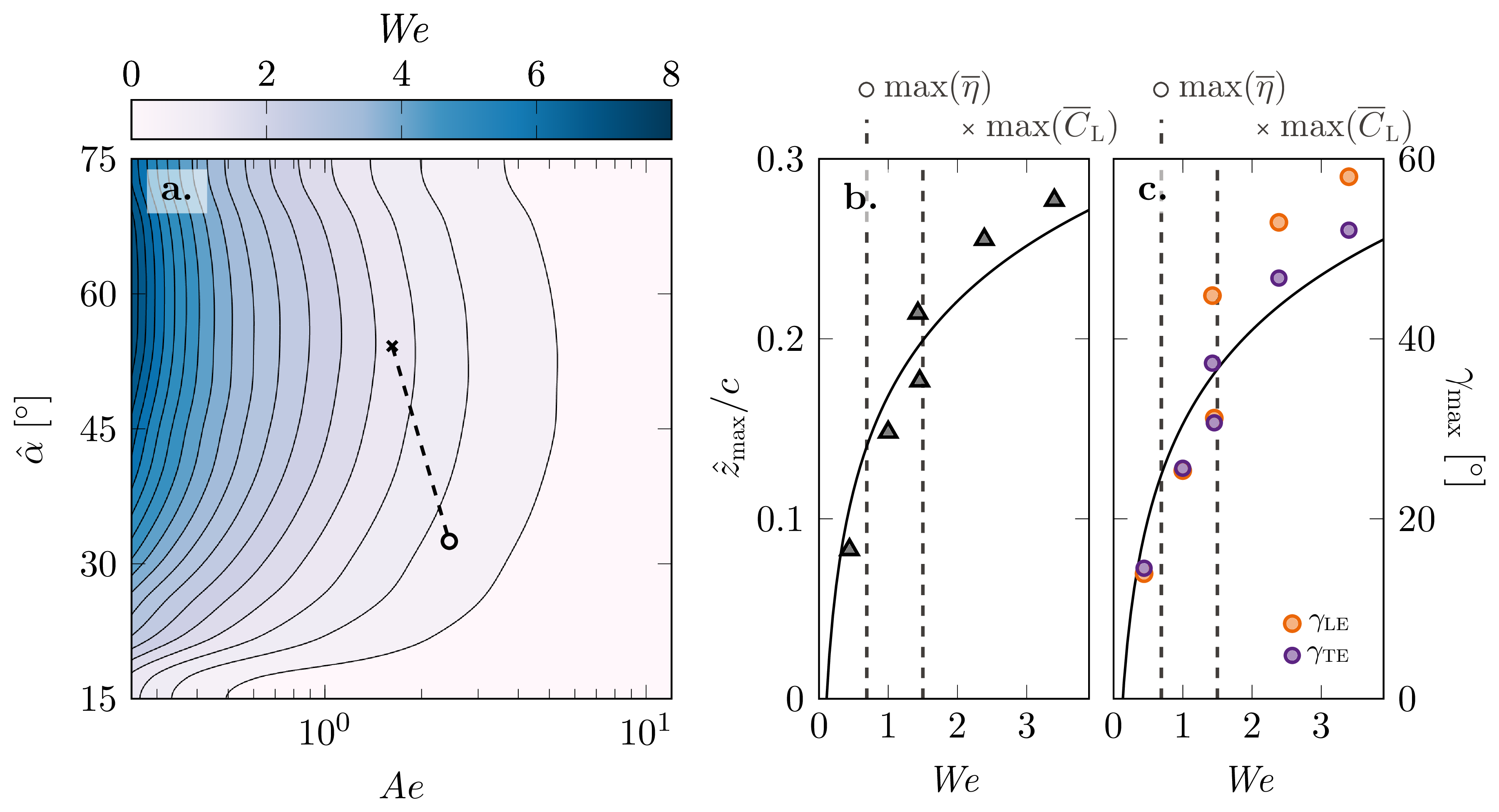}
		\caption{a.~Weber number over aeroelastic number and angle of attack, b.~stroke-maximum camber \kindex{\hat{z}}{max} as a function of the Weber number for $\hat{\alpha} = \ang{55}$, and c.~stroke-maximum leading and trailing edge rotation angles \kindex{\gamma}{LE,max} and \kindex{\gamma}{TE,max} as a function of the Weber number for $\hat{\alpha} = \ang{55}$.}
		\label{fig:WeberVSAeAlpha}
	\end{figure}%
	To quantify the relationship between aerodynamic performance and membrane deformation over the entire experimental parameter range, a Weber number is calculated, $\textit{We} = \kindex{C}{L} / \textit{Ae} = L / (E h R)$ (\cref{eq:weber_number})~\cite{song_aeromechanics_2008, waldman_camber_2017}.
	The Weber number represents the ratio between inertial forces, e.g. lift and drag on the wing, and the membrane tension.
	Previous work on membranes at a steady free-stream velocity and angle of attack demonstrated that the maximum membrane camber $\kindex{\hat{z}}{max}$ increases with increasing Weber number \textit{We} for a pre-strain parameter \kindex{\epsilon}{0} (\cref{eq:weber_number})~\cite{song_aeromechanics_2008, waldman_camber_2017}.
	The trailing edge of the membrane wing in this study can move freely in chord-wise direction and a pre-strain of $\kindex{\epsilon}{0} = 0$ can be assumed.
	The Weber number is presented in \cref{fig:WeberVSAeAlpha}a for all our experiments in the parameter space spanned by \textit{Ae} and $\hat{\alpha}$.
	The two markers indicate the global lift-optimum ($\boldsymbol{\times}$) and efficiency-optimum ($\boldsymbol{\circ}$) found in \cref{fig:ClEtaVsAlphaAe}.
	The Weber number decreases from \textit{We} = \numrange{1.50}{0.69} between the lift and the efficiency optimum.
	With a reduction in Weber number the pressure and lift pulling on the membrane decrease relative to the stiffness of the membrane and less deformation is expected.
	Assuming the membrane camber is a direct function of the Weber number~\cite{song_aeromechanics_2008, waldman_camber_2017}, we can infer the change in maximum membrane camber \kindex{\hat{z}}{max} and deflection angle \kindex{\gamma}{max} from the deformation measurements at $\hat{\alpha} = \ang{55}$ in \cref{fig:WeberVSAeAlpha}b and c.
	For most efficient hovering at \textit{We} = \num{0.69} the camber is predicted to drop below $\kindex{\hat{z}}{max} < 0.15 c$ and limiting $\kindex{\gamma}{max} < \ang{25}$ preventing over-cambering and negative angles of attack at the leading edge.
	The relationship between the increase in the maximum camber $\hat{z}$ and the Weber number in \cref{fig:WeberVSAeAlpha}b agrees qualitatively with experiments conducted for fixed membrane wings without pre-strain~\cite{song_aeromechanics_2008, waldman_camber_2017}.
	Compared to the results under steady flow conditions, the Weber number in our study is one order of magnitude higher.
	This difference in Weber number is attributed to the change to unsteady flapping wing flight and to a difference in the mechanism that causes the membrane deformation.
 	In \cite{song_aeromechanics_2008, waldman_camber_2017}, the membrane deformation is dominated by stretching.
 	In our study it is dominated by bending.\\
	Two distinct regions of maximum lift production and hovering efficiency are identified in the full experimental parameter range.
	The ability to move from high lift production to efficient hovering by modifying the angle of attack or the aeroelastic properties makes the membrane wing platform a promising model for the design of novel micro air vehicles.
	Many natural fliers already modify their wings' angle of attack and stiffness in flight to adjust to different flight situations, for example when performing manoeuvres or to alleviate a gust encounter.
	Different types of flapping wing vehicles have already incorporated angle of attack variation in their designs and progress has been made recently in developing membrane wings with variable stiffness~\cite{curet_aerodynamic_2014, bohnker_control_2019}.
	Our results show that combining both effects, variable stiffness and angle of attack variation, enhances the aerodynamic performance and has the potential of improved control capabilities of micro air vehicles.
	\section{Conclusion}
	We have introduced a novel bio-inspired membrane wing design for systematic investigation of the fluid-structure interactions of flapping membrane wings.
	The wing platform allows for passive cambering of the membrane and for the rotation of the leading and trailing edges with respect to the stroke plane.
	A wide range of kinematic and membrane material parameters were tested on an experimental flapping wing platform and the aerodynamic performance in terms of the stroke-average lift coefficient \kindex{\overline{C}}{L} and hovering efficiency $\overline{\eta}$ was evaluated.
	Additional deformation measurements were conducted over the full stroke duration to capture the temporal evolution of the membrane camber and the orientation of the leading and trailing edges relative to the flow for selected parameter variations.\\
	Across the entire range of angles of attack tested, we found membrane wings that can produce up to \SI{18}{\percent} more lift and reach \SI{16}{\percent} higher lift to power coefficient ratios than the rigid reference wing.
	At the lowest tested angles of attack $\hat{\alpha} < \ang{25}$, the rigid wings perform equally well as the highest performing membrane wings.
	With increasing angles of attack, the rigid wings loose terrain with respect to the membrane wings.
	At the highest angle of attack $\hat{\alpha} = \ang{75}$, the rigid wings have similar performance to the lowest-performing membrane wings.\\
	We computed an aeroelastic number \textit{Ae}, typically used for fixed wings, to characterise the balance between the membrane stiffness and the dynamic pressure on the wing and confirm that it is also suitable to characterise unsteady flapping wings.
	At low \textit{Ae}, the dynamic pressure of the flow on the wing is relatively high compared to the stiffness of the membrane and we observe larger membrane deformations.
	With increasing \textit{Ae}, the effective stiffness of the membrane increases and the membrane deforms less.
	At the highest \textit{Ae}, the membrane behaves the same as a rigid plate.
	For most angles of attack, the stroke-average lift coefficient \kindex{\overline{C}}{L} increases with increasing \textit{Ae} until a maximum is reached at aeroelastic numbers $\textit{Ae}$ ranging from \numrange{1}{2}.
	Further increasing \textit{Ae} beyond the lift optimal aeroelastic conditions leads to a decrease in the lift produced by the membrane wings and their \kindex{\overline{C}}{L} values asymptotically converge to the average lift produced by the rigid wings at the highest \textit{Ae}.\\
	To understand the relationship between effective membrane compliance expressed by the aeroelastic number \textit{Ae} and the aerodynamic force production on the wing, we measured and quantified the membrane deformation throughout the entire wing stroke.
	The stroke-maximum camber \kindex{\hat{z}}{max} grows exponentially with decreasing \textit{Ae} for a given angle of attack.
	As the camber increases, the rotational angle of the leading and trailing edges increase.
	The stroke-average lift coefficient does not increase indefinitely with increasing stroke-maximum camber, but reaches an optimal value near $\kindex{\hat{z}}{max}/c\approx 0.2$ for $\hat{\alpha}=\ang{55}$.
	The lift optimal membrane shape displays two features that enhance the lift production: moderate angles of attack at the leading edge which lead to stall delay and a vertical orientation of the trailing edge which deflects the fluid downwards and enhances the upward reaction force on the membrane.
	For lower \textit{Ae}, the membrane over-cambers and the lift decreases.
	We identify thresholds for over-cambering when either the leading edge angle becomes negative $\kindex{\alpha}{LE,min} < \ang{0}$ or when the trailing edge angle drops below $\kindex{\alpha}{TE,min} < \ang{90}$.\\
	The temporal evolutions of the aerodynamic forces reveal that most of the lift gain of the membrane wings relative to the rigid wings is achieved in the second half of the stroke.
	The power consumption in the second half of the stroke is lower across all tested angles of attack and aeroelastic numbers.
	The membrane camber $\hat{z}$ reaches a maximum around mid-stroke ($t/T = 0.25$), when the stroke velocity $\dot{\phi}$ is highest, and maintains the maximum camber during most of the remainder of the stroke for \textit{Ae} equal or lower than the lift optimal \textit{Ae}.\\
	Across the entire parameter space considered, we identified global maxima for either maximum lift or most efficient hovering at different angles of attack $\hat{\alpha}$ and aeroelastic numbers \textit{Ae}.
	The maximum lift is found at $\textit{Ae} = \num{1.7}$ and $\hat{\alpha} = \ang{55}$ and the most efficient hovering occurs for a higher effective membrane stiffness, $\textit{Ae} = \num{2.4}$, and at a lower angle of attack $\hat{\alpha} = \ang{33}$.
	The aeroelastic numbers we find are in line with previous results reported in literature for flexible hovering wings, membrane energy harvesting applications, and insect species with matching aspect ratio wings.
	In hovering flight, a lower angle of attack typically enhances the lift to drag ratio.
	The same deformation and leading edge rotation angle observed at the lift optimal \textit{Ae} would lead to over-cambering and a negative leading edge angle at the efficiency optimal angle of attack.
	This explains the increase in \textit{Ae} to limit camber and rotation of the leading edge at the efficiency optimum.\\
	Finally, we quantified the relationship between membrane camber, lift production, and the relative stiffness of the membrane wing using the Weber number \textit{We}.
	For the lift and the efficiency optimal cases, the Weber number decreases from \textit{We} = \numrange{1.50}{0.69} suggesting a reduction in membrane tension and membrane deformation.
	With a reduction in Weber number, the pressure and lift pulling on the membrane decrease relative to the stiffness of the membrane.
	The lower \textit{We} is associated with a decrease in membrane camber which limits the leading edge rotation angle $\kindex{\gamma}{max} < \ang{25}$ preventing over-cambering and negative angles of attack at the leading edge.\\
	The ability to move from high lift production to efficient hovering by modifying the angle of attack or the aeroelastic properties makes the membrane wing platform a promising model for the design of novel micro air vehicles.
	Our results show that combining both effects, variable stiffness and angle of attack variation, enhances the aerodynamic performance of membrane wings and has the potential to improve control capabilities of micro air vehicles.
	\section*{Author Contributions}%
	%
	%
	\textbf{AG:} Conceptualisation, Investigation, Formal analysis, Writing - original draft;
 	\textbf{JR:} Resources, Visualisation;
	\textbf{EU:} Conceptualisation, Resources;
	\textbf{KM:} Funding acquisition, Supervision, Writing - Review \& Editing.\\
	\section*{Acknowledgments}%
	This work was supported by the Swiss National Science Foundation under grant number 200021\_175792
	\section{References}%
	\bibliography{ms}
	\bibliographystyle{ieeetr}

	\appendix
	\section{Convergence of the mean force coefficient}
	\label{sec:force_average_convergence}
\begin{figure}
		\centering
		\includegraphics[]{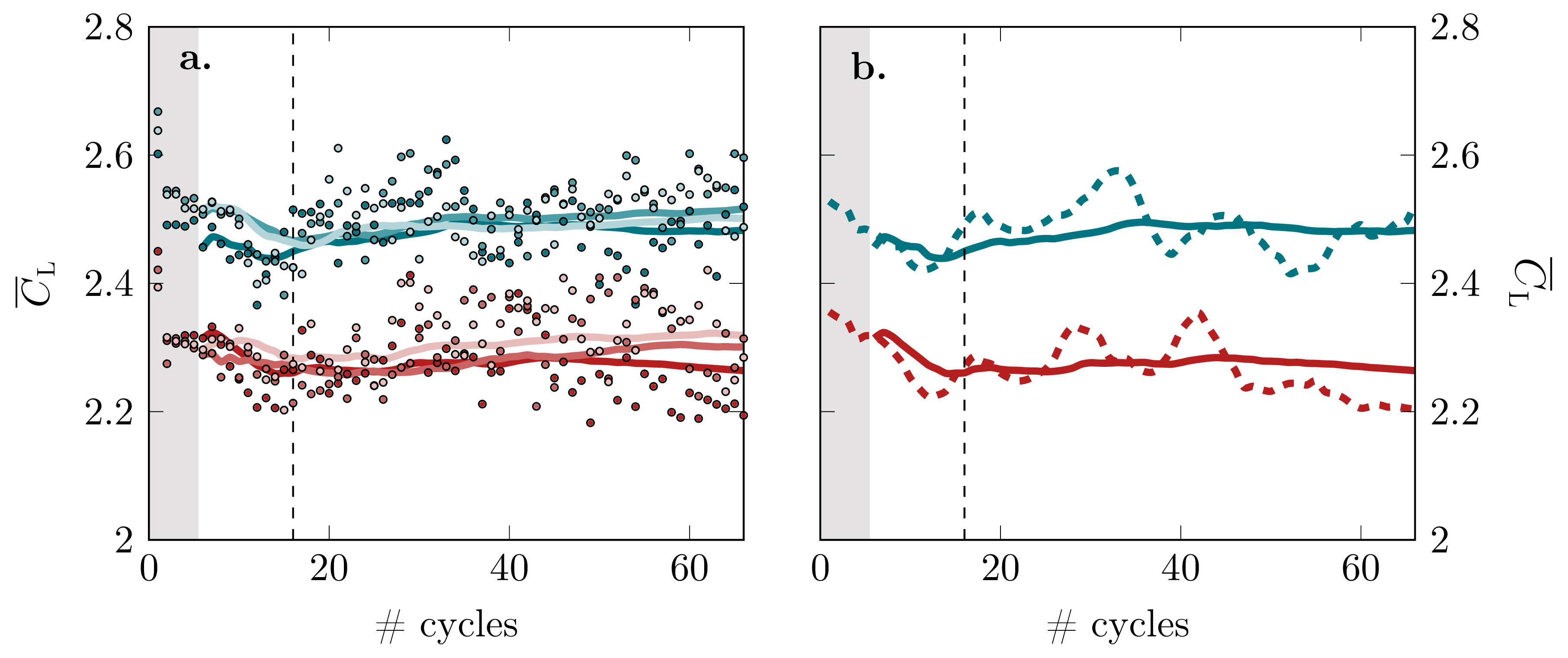}
		\caption{Average lift coefficient recorded over 66 cycles for two different frequencies.
The different colours refer to different cases and the different shades indicate the different repetitions of the same parameter frequency.
The markers indicate the cycle-average lift coefficient for the individual cycles and the solid lines represent the cumulative average lift starting from the sixth cycle.}
		\label{fig:mCL_manyCycles}
	\end{figure}

	Large-scale recirculation of the fluid within the tank due to the down-wash created by the wing can affect the force measurements when repeating many flapping cycles in a closed vessel.
	In the following section we present a quantitative study to determine the influence of the tank confinement on the average force data.
	\Cref{fig:mCL_manyCycles}a summarises the experimental results for two different cases with each three repetitions conducted over $66$ cycles.
	This is equivalent to an experimental time of $5$ minutes or more than $4$ times the number of cycles of the main study for the higher lift producing case with a frequency of $f = \SI{0.225}{\hertz}$.
	The markers show the cycle-average lift coefficient for the individual cycles and the solid lines represent the cumulative average lift starting from the sixth cycle.
	The grey area indicates the first five cycles which are being removed from the averaging and the vertical dashed line marks the end of $16$ cycles which is the number of cycles considered in the main study.
	\Cref{fig:mCL_manyCycles}b shows a close-up view with only one out of the three cases for visual clarity.
	Here, the dashed lines show a moving average over $5$ cycles to highlight potential long term drift of the forces due to recirculation in the tank.
	Even though we see some variation in the moving average lift, the cumulative average values do not change substantially after the first $16$ cycles.
	The largest relative observed drift in the cumulative mean is found for the top case in \cref{fig:mCL_manyCycles}b which varies from $\kindex{\overline{C}}{L,n=16} = 2.449$ to $\kindex{\overline{C}}{L,n=37} = 2.496$ which corresponds to a \SI{1.9}{\percent} increase.
	\section{Wing inertia}
	\label{sec:wing_inertia}
\begin{figure}
		\centering
		\includegraphics[]{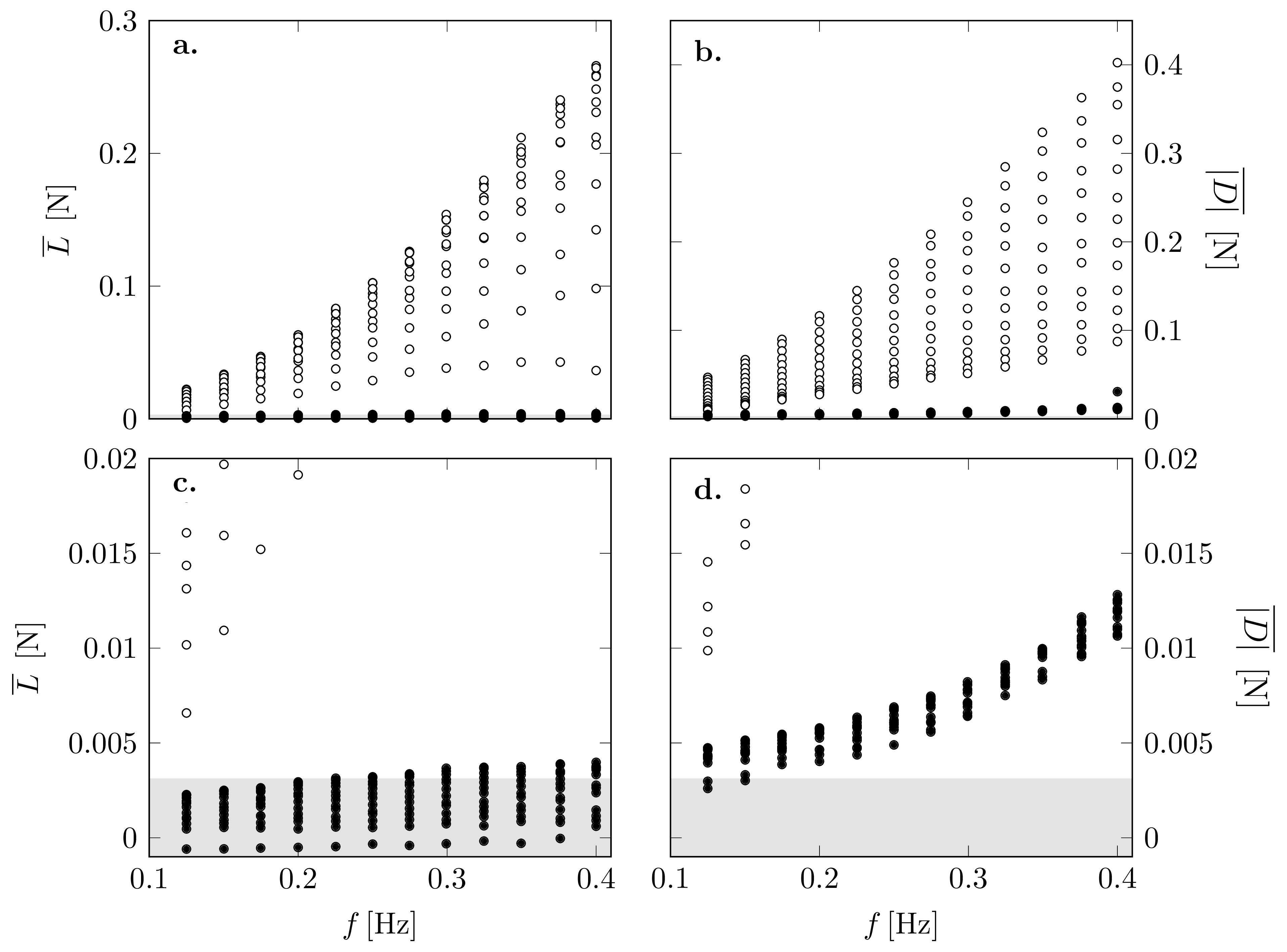}
		\caption{Dimensional average lift force $\overline{L}$ and average of the absolute drag force $\overline{\left| D \right|}$ for the membrane wing with thickness $t = \SI{1.4}{\milli\metre}$.
The open markers show the stroke-average forces in water and the filled markers show the results in air.
		}
		\label{fig:wing_inertia_quantification}
	\end{figure}

	Depending on the mass ratio between the wing and the fluid, inertial forces can be of the same order of magnitude as the aerodynamic forces for flapping wings in air~\cite{gopalakrishnan_effect_2010}.
	Our experiments are conducted in water which allows us to achieve low density ratios between the membrane wing and the fluid $\kindex{\rho}{m} / \kindex{\rho}{water} = 1.20$.
	We conducted additional experiments in air where the inertial forces dominate the measurement loads.
	In \cref{fig:wing_inertia_quantification} we present the dimensional average lift force $\overline{L}$ and the average of the absolute drag force $\overline{\left| D \right|}$ for the heaviest membrane wing with thickness $t = \SI{1.4}{\milli\metre}$, which has the highest contribution of the inertial forces to the total measured force.
	The open markers show the stroke-average forces in water and the filled markers show the results in air.
	The first row of images shows an overview of the entire measurement set and the second row of images shows a zoomed in version on the experiments in air.
	The grey areas highlight the force transducer resolution of \SI{3.13}{\milli\newton} around zero.
	The inertial forces recorded in air increase with increasing frequency but are close to or even below the resolution of the load cell.
	For our set-up, the inertial forces are much lower than the fluid mechanics forces and the former are deemed negligible.
\end{document}